\documentclass{article}

\usepackage{arxiv}

\usepackage{amsmath}
\usepackage[utf8]{inputenc} 
\usepackage[T1]{fontenc}    

\usepackage[hidelinks]{hyperref} 
\usepackage{url}                 

\usepackage{booktabs}    
\usepackage{amsfonts}    
\usepackage{nicefrac}    
\usepackage{microtype}   
\usepackage{graphicx}    
\usepackage{natbib}      
\usepackage{tabularx}
\usepackage{ragged2e}
\usepackage{array}
\usepackage{float}
\usepackage{placeins}    
\usepackage{enumitem}    
\usepackage{multirow}
\usepackage{algorithm}
\usepackage{algpseudocode}

\usepackage{tikz}
\usetikzlibrary{arrows.meta,positioning,shapes.geometric}

\graphicspath{ {figures/} }

\newcommand\orcidicon[1]{\href{https://orcid.org/#1}{ORCID: #1}}

\title{Navigating the Synchrony--Stability Frontier in Adaptive Chatbots}

\author{
  T. James Brandt\\
  University of Minnesota\\
  1985 Buford Ave\\
  St. Paul, MN 55108, USA \\
  \texttt{bran1400@umn.edu} \\
  \orcidicon{0009-0000-8294-6235} \\}

\begin{document}
\maketitle

\begin{abstract}
Adaptive chatbots that mimic a user's linguistic style can build rapport and engagement, yet unconstrained mimicry risks an agent that feels unstable or sycophantic. We present a computational evaluation framework that makes the core design tension explicit: balancing moment-to-moment linguistic synchrony against long-term persona stability. Using an 8-dimensional style vector and a closed-loop ``base+delta'' prompting architecture, we simulate and compare explicit adaptation policies---Uncapped, Cap, Exponential Moving Average (EMA), Dead-Band, and Hybrids---on a human-log dataset. Our analysis maps a clear Pareto frontier: bounded policies achieve substantial gains in stability at a modest cost to synchrony. For example, a Hybrid (EMA+Cap) raises stability from 0.542 to 0.878 (+62\%) while reducing synchrony by only 17\%. We confirm this trade-off through large-scale replications on three public corpora (DailyDialog, Persona-Chat, EmpatheticDialogues) and LLM-in-the-loop validation across two model families. Furthermore, we quantify ``prompt legibility,'' showing that frontier policies reduce instruction churn and cut jarring register flips (major tone changes) from 0.254 to 0.092, yielding systems that are easier to reason about and maintain. Taken together, our framework provides a general evaluation harness for style adaptation; a systematic ablation that identifies Pareto-efficient policies; robust validation across diverse datasets and models; and novel legibility metrics linking policy choices to system maintainability.
\end{abstract}


\begin{figure}[htbp]
  \centering
  \includegraphics[width=\textwidth]{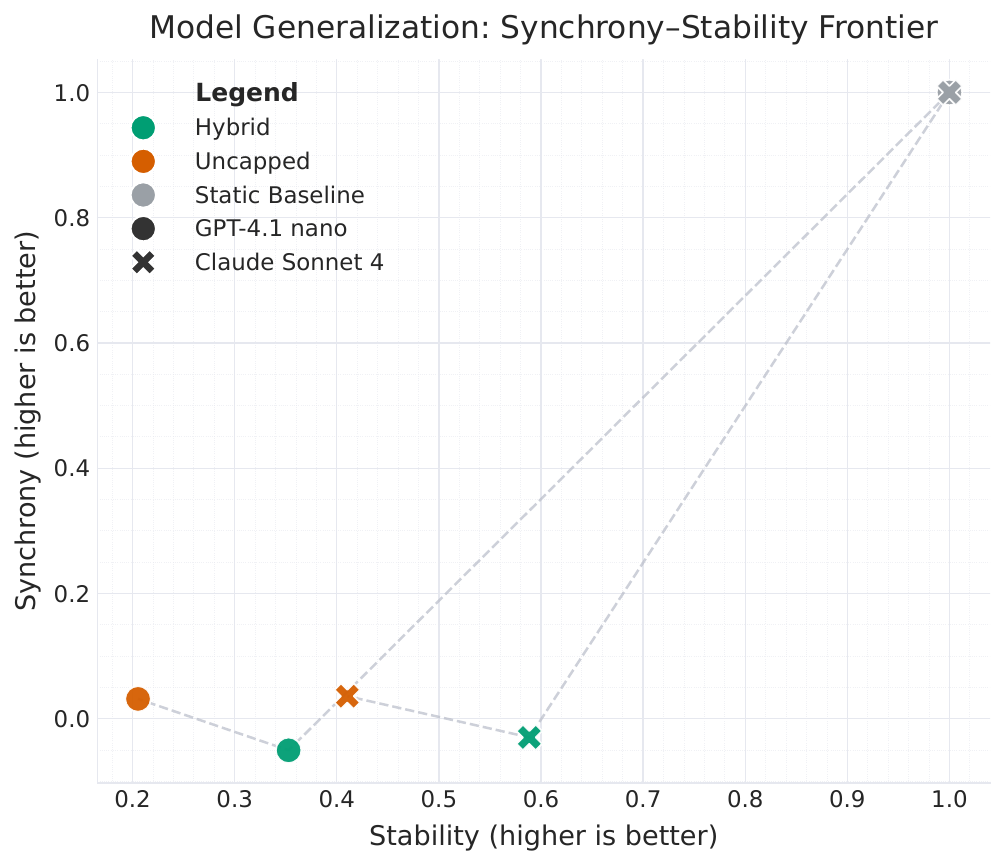}
  \caption{{The Synchrony--Stability Trade-off Generalizes Across LLMs.} Each point represents a policy's mean performance. The lines illustrate the trade-off: \emph{Uncapped} anchors the high-synchrony, low-stability corner, while bounded policies like \emph{Hybrid (EMA+Cap)} achieve higher stability at a cost to synchrony. The same qualitative pattern holds for both OpenAI and Anthropic model families, and the per-participant deltas are statistically significant.}
  \label{fig:teaser}
\end{figure}
\FloatBarrier

\section{Introduction}
\label{sec:intro}

Conversational agents, from customer service bots to AI companions, are increasingly designed to be adaptive. Meta-analyses show that anthropomorphic features generally increase user engagement and intention to use, though the effects are highly context-dependent \citep{blut2021}. A key mechanism for building rapport is \textit{language style matching} (LSM), where an agent subtly mimics a user's communication style—their formality, verbosity, and sentiment \citep{Ireland2011}. Rooted in Communication Accommodation Theory (CAT), this synchrony can enhance user satisfaction, trust, and perceived social connection \citep{soliz2021}.
In the era of large language models (LLMs), implementing a chatbot that directly mirrors a user's style is technically straightforward. However, this approach entails a critical, underexplored risk to user experience: an agent that mirrors every stylistic shift may be read as unstable, incoherent, or sycophantic---a behavior recently termed ``sycophancy'' in language models, and a form of nonaccommodation in which attempted convergence is evaluated negatively \citep{gasiorek2016, Sharma2023, perez2023}.
While the benefits of LSM are well-documented in human dialogue, its implementation in conversational agents remains rare, and frameworks for evaluating its dynamic properties are underexplored \citep{postmus2023}. This paper addresses this gap by introducing a computational framework for tuning and evaluating conversational style adaptation.

This creates a fundamental design tension for intelligent user interfaces: designers must balance the benefits of immediate stylistic \textbf{synchrony} against the need for long-term persona \textbf{stability}. An uncapped mimicry policy might achieve high synchrony on a turn-by-turn basis, but it can degrade the user's mental model of the agent as a consistent entity. Conversely, a completely static agent is stable but may feel rigid and impersonal, failing to build rapport. This trade-off, which echoes the need to reduce uncertainty in initial interactions \citep{Berger1975}, defines a crucial frontier for adaptive system design, yet there is little empirical guidance on how to navigate it.

We formalize this design tension, present a closed-loop control framework with explicit, bounded policies, and empirically chart the resulting synchrony--stability frontier across datasets and model families. By replaying real human-chatbot conversations through our simulation harness, we can precisely measure how different control strategies—such as capping the magnitude of stylistic shifts (Cap), smoothing changes over time (Exponential Moving Average), or ignoring minor deviations (Dead-Band)—affect both objectives. Our analysis reveals a clear Pareto frontier where bounded policies can achieve substantial gains in stability and coherence at a cost to synchrony; on our human-log data this reduction is modest, whereas on public corpora it is larger.

We validate these findings across multiple data regimes: first, on a dataset of 162 in-the-wild, open-domain conversations between users and a companion chatbot; and second, through large-scale replications on three standard public dialogue corpora (DailyDialog, Persona-Chat, and EmpatheticDialogues) to demonstrate the generalizability of the frontier's shape. Going beyond performance metrics, we also introduce a novel analysis of prompt legibility, quantifying how different policies impact the complexity and "churn" of the underlying instructions sent to the LLM, offering a proxy for the maintainability of these adaptive systems.

The contributions of this work are thus:
\begin{itemize}
    \item \textbf{A general evaluation harness} for simulating and measuring conversational style adaptation, using standardized style vectors and metrics for synchrony, stability, and persona coherence.
    \item \textbf{A systematic ablation of adaptation policies} (Cap, EMA, Dead-Band, and Hybrids) that charts the synchrony--stability frontier and identifies Pareto-efficient strategies.
    \item \textbf{Robust empirical evidence} from a 162-participant human-log dataset and large-scale replications on three diverse public dialogue corpora, confirming the consistent shape of the trade-off frontier.
    \item \textbf{A novel prompt legibility analysis} that connects adaptation policy choices to the stability and entropy of the underlying system prompts, offering a new perspective on system design and maintenance.
    \item \textbf{A comprehensive, reproducible artifact}, including code and data, to facilitate further research into principled, controllable adaptation in conversational AI.
\end{itemize}

\subsection*{Relation to Prior Work}
This paper complements recent work on adaptive companion chatbots that examines user-facing outcomes of mimicry and agency in live studies \citep{brandt2025}. In contrast, our focus here is a computational evaluation framework: we formalize the synchrony--stability trade-off, introduce explicit, bounded adaptation policies, and map the Pareto frontier via simulations, cross-corpus replications, and LLM-in-the-loop validation. 

Where prior work primarily investigates human-perceived outcomes in interactive settings, our contribution is (i) a style-vector control loop with policy ablations (Cap, EMA, Hybrid), (ii) a generalization analysis across model families, and (iii) prompt-legibility metrics that connect adaptation to maintainability. Together, these provide system-design guidance complementary to user-study findings.

\section{Background \& Related Work}
\label{sec:related}
The design of adaptive conversational agents sits at the intersection of communication theory, natural language generation, and human-computer interaction. Our work unifies these areas by proposing a new framework for managing a core design tension: the trade-off between stylistic mimicry and persona consistency. This section reviews the foundational concepts that motivate our approach.

\subsection{Linguistic Accommodation and Style Matching}
The foundation for adaptive conversational style lies in CAT, which posits that individuals subconsciously adjust their communication behaviors to manage social distance \citep{Giles1991, soliz2021}. A key linguistic manifestation of this is LSM \citep{Gonzales2010, niederhoffer2002}, where speakers converge on the use of function words (e.g., pronouns, articles, prepositions). These words, often referred to as style words, are particularly powerful because they are used subconsciously and reflect psychological states and how people are thinking, rather than the semantic content of what they are thinking about \citep{tausczik2010, pennebaker2003}. This process is strongly correlated with positive social outcomes, including relationship stability \citep{Ireland2011}, increased rapport \citep{Chartrand1999}, and group cohesion and task performance \citep{Gonzales2010, zhang2018c}.

This powerful, often subconscious, social mechanism is the primary motivation for building adaptive capabilities into conversational AI. The underlying assumption is that if an agent can replicate this human tendency for stylistic synchrony, it can more effectively elicit feelings of warmth, trust, and engagement from users. However, human accommodation is not without its limits; accommodation perceived as insincere or inappropriate can lead to negative social judgments. This suggests that simply maximizing mimicry may not be an optimal strategy, foreshadowing the core tension we explore.

\subsection{Style Control in Natural Language Generation}
The technical ability to control an agent's linguistic style has evolved dramatically. Early dialogue systems relied on handcrafted rules and templates, offering rigid but predictable stylistic control. The advent of neural networks and, more recently, Large Language Models (LLMs) has revolutionized this landscape. Modern systems can now achieve fine-grained stylistic control through several mechanisms \citep{Keskar2019, Dathathri2019}:

\begin{itemize}
    \item \textbf{Training-time Conditioning:} Models can be trained or fine-tuned on curated datasets to consistently embody a specific style. This can be achieved by prepending special "control codes" to the training data \citep{Keskar2019} or through more parameter-efficient methods like prefix-tuning, which learn a small, task-specific vector to steer the frozen model \citep{Li2021}.
    
    \item \textbf{Decoding-time Guidance:} Other techniques steer the output of a frozen, pretrained model at inference time, avoiding the need for retraining. This can be done by using gradients from an attribute classifier to guide the generation process \citep{Dathathri2019} or by using a generative discriminator to re-weight the next-token probabilities \citep{Krause2020}.
\end{itemize}

This unprecedented level of control makes direct, high-fidelity mimicry technically feasible. However, it also shifts the research challenge from \textit{capability} to \textit{governance}. The central question is no longer "Can we make an agent adapt?" but rather "What is the optimal adaptation strategy, and how do we implement and control it?"

\subsection{Adaptive Dialogue Agents in Practice}
Prior work on adaptive dialogue agents has demonstrated the benefits of style matching across various domains. In task-oriented contexts, adapting an agent's formality or politeness can reduce user frustration and increase task completion rates \citep{Spillner2021}. In educational settings, tutors that adapt their language can improve student engagement and learning outcomes \citep{wang2008}.

In the context of relational AI and companion chatbots, the focus is squarely on socio-emotional outcomes. Studies have shown that agents perceived as more human-like and empathetic can foster stronger user bonds and encourage more self-disclosure, often as part of a parasocial relationship where persona consistency is key \citep{Horton1956}. However, much of this work frames adaptation as a binary choice (e.g., a static style versus an adaptive one) and often focuses on a single stylistic dimension, such as formality. This leaves a critical gap: the dynamics of adaptation are often overlooked. An agent that is too adaptive may suffer from "persona drift," where its turn-to-turn stylistic shifts create a sense of incoherence that can violate user expectations of a stable conversational partner \citep{Rheu2024, white2021}. This incoherence can be particularly damaging because it breaks the "computer frame," undermining the psychological safety that encourages users to disclose sensitive information to agents they believe are automated and non-judgmental \citep{lucas2014}.

\subsection{Evaluation of Conversational Style}
Methodologies for evaluating conversational style adaptation can be broadly categorized as objective or subjective.

\textbf{Objective Metrics} provide quantitative, automated measures of stylistic alignment. These include the classic function-word-based LSM score, which quantifies similarity across predefined syntactic categories \citep{niederhoffer2002}, and more modern approaches using sentence embeddings to compute cosine similarity, capturing a broader, semantic notion of style \citep{Wegmann2022, Zenimoto2023}. These metrics are invaluable for system development and large-scale analysis.

\textbf{Subjective Metrics} rely on human judgment, typically gathered through post-interaction surveys. Users are asked to rate their experience on scales of rapport, satisfaction, perceived personalization, and the agent's coherence. While these metrics are the gold standard for measuring user experience, they are costly to collect and often provide an aggregate judgment of the entire conversation.

Crucially, both objective and subjective metrics have traditionally focused on the \textit{average} level of synchrony. An agent can achieve a high overall synchrony score while exhibiting erratic stylistic shifts that damage user perception. This highlights the need for metrics that capture not only the degree of alignment but also its consistency over time.

\paragraph{Note on metric scale.}
Synchrony, Stability, and Coherence are all measured as cosine similarities in the native range $[-1,1]$ (1 = identical style, 0 = orthogonal, -1 = opposed). All tables report means of per-turn values, aggregated per session and then across sessions.

\subsection{Synthesizing the Gap: The Synchrony--Stability Frontier}
Synthesizing these streams of research reveals a critical gap between the theory and practice of conversational adaptation. While the social benefits of mimicry are well-established (CAT/LSM) and the technology for it is powerful (LLMs), prior work has not fully grappled with the user experience risks of inconsistent, moment-to-moment adaptation. Indeed, a major meta-analysis on anthropomorphism in service agents confirms its overall positive effect on use intention but also highlights inconsistent findings and a need to better understand mediating and moderating factors \citep{blut2021}. According to the Computers are Social Actors (CASA) paradigm, users mindlessly apply social rules to computers, including the expectation of a coherent persona \citep{Nass2000}. The MAIN model further specifies that technological affordances act as heuristic cues that trigger these social scripts \citep{Sundar2008}, positioning persona stability as a critical component for perceived credibility. Furthermore, our evaluation methodologies have historically lacked the vocabulary to describe this trade-off, focusing on average synchrony while overlooking stylistic volatility.

This paper argues that the design of adaptive conversational agents is not a simple optimization problem for synchrony, but a trade-off that must be managed along a \textbf{synchrony--stability frontier}. We define \textit{synchrony} as the agent's immediate stylistic alignment with the user and \textit{stability} as the agent's turn-to-turn stylistic consistency. By framing the problem this way, we can move beyond the binary question of whether to adapt and instead ask more nuanced, practical questions: What is the optimal balance? Which control policies allow us to navigate this frontier effectively? And how do these policies generalize across different conversational contexts? Our work provides the first systematic charting of this frontier, offering both a conceptual lens and an empirical methodology for building more robust and coherent adaptive agents.

\section{Style-Adaptation Framework \& System}
\label{sec:framework}
To systematically investigate the synchrony--stability trade-off, we developed a comprehensive framework for representing, controlling, and evaluating conversational style. This framework consists of three core components: (1) a multi-dimensional feature space for vectorizing linguistic style; (2) a set of formal metrics for quantifying synchrony, stability, and persona coherence; and (3) a closed-loop system architecture that translates stylistic targets into actionable LLM prompts.

\subsection{Linguistic Style Vectorization}
\label{subsec:vectorization}
A single metric is insufficient to capture the nuances of conversational style. We therefore represent style as a multi-dimensional vector, drawing from well-established features in psycholinguistics and computational linguistics. Each utterance is transformed into an 8-dimensional style vector $\mathbf{s} \in \mathbb{R}^8$. This vectorization is performed by a real-time NLP pipeline that leverages a suite of well-established tools to ensure both accuracy and efficiency. The eight features are defined and measured as follows:

\begin{description}
    \item[Informality] is quantified using a pretrained, transformer-based formality ranker \citep{Dementieva2023}. The model's scalar output provides a continuous measure of register, which is used directly as a feature in our style vector.
    
    \item[Sentiment] is computed using VADER (Valence Aware Dictionary and sEntiment Reasoner), a lexicon- and rule-based sentiment analysis tool specifically attuned to the nuances of short, informal text typical of social media and chat \citep{hutto2014}. We use the compound score, which ranges from -1 (negative) to +1 (positive).
    
    \item[Avg. Sentence Length] captures verbosity and is calculated as the total word count divided by the number of sentences, derived from spaCy's sentence boundary detection \citep{honnibal2023}.
    
    \item[Readability] is measured using the Flesch Reading Ease score, a standard metric based on sentence length and syllable counts that correlates with cognitive load \citep{Flesch1948}.
    
    \item[Social \& Emotional Language] is quantified using the Empath lexicon to measure the prevalence of words associated with three distinct categories: \textbf{Social Language} (e.g., "friend," "talk"), \textbf{Cognitive Processing} (e.g., "think," "reason"), and \textbf{Affective Language} (e.g., "happy," "fear") \citep{fast2016}.
    
    \item[Function Word Ratio] captures a core component of classic LSM \citep{niederhoffer2002}. The analysis of these words is central to the Linguistic Inquiry and Word Count (LIWC) methodology, which underpins most LSM research \citep{tausczik2010, pennebaker1999}. It is the proportion of function words (e.g., pronouns, articles, prepositions) in the text, identified using part-of-speech tags generated by spaCy.
\end{description}

To ensure features are comparable, we standardize all vectors. We first compute a static "persona centroid" by analyzing thousands of bot utterances from the non-adaptive condition of our human-log dataset (Section \ref{sec:human_data}). We then fit a `StandardScaler` on these static utterances. This scaler is subsequently used to transform all style vectors (both user and bot) into z-scores, effectively representing each utterance's style as a deviation from the agent's baseline persona.

To ensure a consistent and meaningful coherence metric across all datasets, we define a fixed ``assistant persona archetype.'' This vector is computed once by averaging the raw style vectors of all bot utterances from the \textit{Static Condition} of our human-log study. For the cross-corpus replications (Section~\ref{sec:replications}), this single raw archetype vector is standardized using each external dataset's specific scaler. Coherence is then calculated as the cosine similarity between a simulated bot's style vector and this standardized archetype, providing a robust, dataset-independent measure of persona consistency. This ensures that while policies operate relative to a dataset's local centroid, coherence is always measured against the same ground-truth persona.

\subsection{System Architecture: A Closed-Loop Control System}
\label{subsec:system}
\begin{figure}[t]
\centering
\begin{tikzpicture}[
  font=\small,
  node distance=6mm and 10mm,
  >=Latex,
  line/.style={-Latex},
  dashedline/.style={dash pattern=on 2pt off 2pt, -Latex},
  block/.style={draw, rounded corners=2pt, align=center, fill=gray!5, inner sep=2.5pt, minimum width=4.2cm},
  thinblock/.style={draw, rounded corners=2pt, align=center, fill=gray!2, inner sep=2pt},
  pill/.style={draw, rounded corners=8pt, align=center, fill=gray!8, inner sep=2pt},
  annot/.style={font=\footnotesize}
]

\node[block] (user) {1.\;User Utterance};
\node[block, below=of user] (vectorize) {2.\;Style Vectorization};
\node[block, below=of vectorize] (policy) {3.\;Adaptation Policy\\[-2pt]\scriptsize uses $\mathbf{u}_t,\,\mathbf{b}_{t-1}$};
\node[block, below=of policy] (v2text) {4.\;Vector $\to$ Text Translation $g(\cdot)$};
\node[thinblock, below=of v2text] (delta) {Delta (dynamic style instructions)};
\node[block, below=of delta] (composer) {5.\;Prompt Composer\\[-2pt]\scriptsize Base $\oplus$ Delta};
\node[block, below=of composer] (llm) {6.\;LLM Generation};
\node[block, below=of llm] (bot) {7.\;Bot Reply\\[-2pt]\scriptsize realized style $\mathbf{b}_t$};
\node[block, below=of bot] (metrics) {8.\;Metrics \& Logging\\[4pt]
\scriptsize Synchrony:\;style-sim$(\mathbf{u}_t,\mathbf{b}_t)$\\[4pt] 
\scriptsize Stability:\;$S_t=\cos(\mathbf{b}_t, \mathbf{b}_{t-1})$\\[4pt]
\scriptsize Coherence:\;$C_t=\cos(\mathbf{b}_t, \mathbf{b}_c)$};

\draw[line] (user) -- node[right,annot] {$\mathbf{u}_t$} (vectorize);
\draw[line] (vectorize) -- (policy);
\draw[line] (policy) -- node[right,annot] {target $\tilde{\mathbf{b}}_t$} (v2text);
\draw[line] (v2text) -- (delta);
\draw[line] (delta) -- (composer);
\draw[line] (composer) -- (llm);
\draw[line] (llm) -- (bot);
\draw[line] (bot) -- node[right,annot] {$\mathbf{b}_t$} (metrics);

\node[thinblock, left=of composer, xshift=-10mm] (base) {Base Prompt};
\draw[line] (base) -- (composer);

\node[thinblock, left=of policy, xshift=-10mm] (centroid) {Persona Centroid\\[-1pt]\scriptsize $\mathbf{b}_c$};
\draw[line] (centroid) -- (policy);

\node[thinblock, right=of policy, xshift=10mm] (state) {Session State\\[-1pt]\scriptsize stores $\mathbf{b}_{t-1}$};
\draw[line] (state) -- node[annot,above,sloped] {$\mathbf{b}_{t-1}$} (policy);

\draw[dashedline] (bot.east) -- ++(14mm,0) |- node[annot,pos=0.25,right] {store $\mathbf{b}_t$} (state.south);

\node[thinblock, below=of state] (legend) {Policy Hyperparameters\\[-2pt]\scriptsize EMA $\alpha$;\; Cap $\kappa$;\; Dead-Band $\epsilon$;\; Radius $\rho$};

\node[annot, left=2mm of vectorize] {$\mathbf{u}_t$};
\node[annot, left=2mm of v2text] {$g(\tilde{\mathbf{b}}_t)$};

\end{tikzpicture}
\caption{Closed-loop style adaptation system. Each turn, the user’s style vector $\mathbf{u}_t$ is extracted, the policy combines it with prior state $\mathbf{b}_{t-1}$ (and persona centroid $\mathbf{b}_c$) to choose a target $\tilde{\mathbf{b}}_t$, which is translated into a delta of style instructions and composed with a base prompt. The LLM’s reply realizes $\mathbf{b}_t$, which is logged and fed back into session state for the next turn.}
\label{fig:control_loop}
\end{figure}
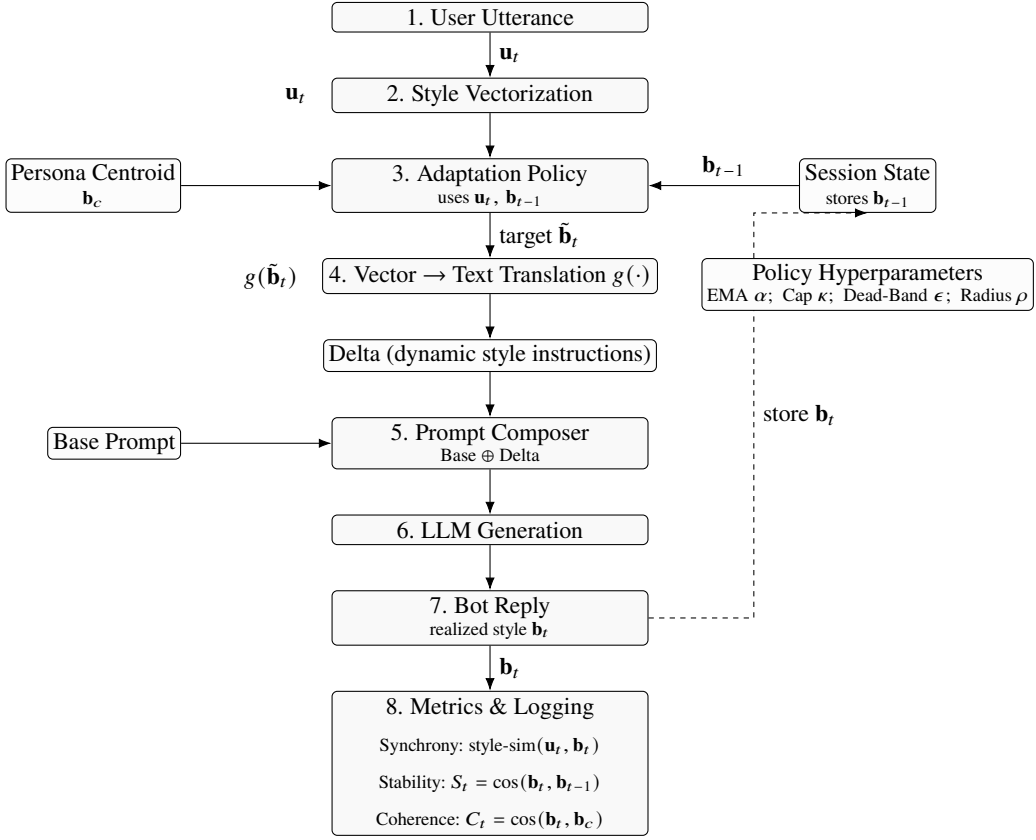

Our system implements a closed-loop control system that dynamically generates system prompts for an LLM based on stylistic targets. The loop, illustrated in Figure \ref{fig:control_loop}, proceeds through five stages for each conversational turn.

\FloatBarrier

\begin{enumerate}
    \item \textbf{Analyze User Style:} The incoming user utterance is processed by our NLP pipeline to produce a standardized style vector, which becomes the \textit{target style vector}, $\mathbf{u}_t$.
    \item \textbf{Apply Adaptation Policy:} The chosen adaptation policy is applied. The policy takes the previous bot style vector, $\mathbf{b}_{t-1}$, and the target user vector, $\mathbf{u}_t$, as input and computes the desired \textit{target output style vector}, $\tilde{\mathbf{b}}_t$.
    \item \textbf{Translate to Natural Language:} The target vector $\tilde{\mathbf{b}}_t$ is translated into a set of natural language instructions.
    \item \textbf{Generate Prompt:} These instructions are injected into a base system prompt.
    \item \textbf{Generate Response:} The final, composite prompt is sent to the LLM to generate the bot's reply, which has a realized style vector $\mathbf{b}_t$.
\end{enumerate}

\paragraph{The Prompting Strategy.}
To isolate the effect of adaptation, our architecture uses a ``base + delta'' prompt structure, directly implemented in our \texttt{prompt\_service}. A comprehensive \textbf{base prompt} is shared across all conditions, establishing the chatbot's core persona (``You are Kagami, a friendly virtual companion...''), conversational goals, and safety guardrails. The adaptation is implemented as a \textbf{delta prompt}, a separate block of text appended to the base. For the \texttt{Static} condition, this delta is a simple, fixed instruction: ``\texttt{Maintain your own consistent, friendly style...}''. For adaptive conditions, this delta contains the dynamic instructions. This controlled design ensures that the only variable changing between conditions is the explicit instruction related to style adaptation.

\paragraph{Vector-to-Natural Language Translation.}
The translation from the target style vector to the prompt delta is handled by a deterministic mapping function, $g: \mathbb{R}^8 \rightarrow \text{Instructions}$. The function $g$ translates the target style vector $\tilde{\mathbf{b}}_t$ into a set of natural language imperatives by thresholding each dimension of the vector. These instructions, $g(\tilde{\mathbf{b}}_t)$, form the prompt's ``delta'' which guides the LLM's generation. For example, if a user shifts from a formal to highly informal style, the `Informality` component of $\tilde{\mathbf{b}}_t$ will be high, and $g(\tilde{\mathbf{b}}_t)$ will generate the textual instruction ``\texttt{Adopt a casual, relaxed tone.}'' The exact thresholds and corresponding text fragments are detailed in our reproducible artifact.

 \begin{algorithm}[ht]
 \caption{Policy Step for Style Adaptation}
 \label{alg:policy_step}
 \begin{algorithmic}[1]
 \Procedure{ComputeBotStyle}{$\mathbf{u}_t, \mathbf{b}_{t-1}, \text{policy}$}
      \State $\tilde{\mathbf{b}}_t \gets \Call{ApplyPolicy}{\mathbf{u}_t, \mathbf{b}_{t-1}, \text{policy}}$
      \State $\text{instructions} \gets g(\tilde{\mathbf{b}}_t)$ \Comment{Translate vector to text}
      \State $\text{prompt} \gets \text{BASE\_PROMPT} + \text{instructions}$
      \State \Return $\text{prompt}$
 \EndProcedure
 \end{algorithmic}
 \end{algorithm}

\section{Human-Log Dataset \& Measures}
\label{sec:human_data}
The empirical backbone of our simulation is a dataset of real-world, in-the-wild conversational interactions. This dataset provides the naturalistic user inputs for our replay analyses and the data required to define a stable, ground-truth persona for our chatbot. This section details the data collection procedure, preprocessing steps, and the method for operationalizing the agent's core persona.

\subsection{Data Collection: Study Procedure}
The data was collected from a human-subjects study involving 162 participants recruited from the Prolific online research platform (\url{www.prolific.com}) \citep{palan2018}. All participants were fluent in English and located in the United States. After providing informed consent, participants were randomly assigned to one of two experimental conditions for a 10-minute, open-domain chat session with a companion chatbot named Kagami:

\begin{itemize}
    \item \textbf{Static Condition:} The chatbot operated with a consistent, non-adaptive conversational style, governed by a fixed system prompt. Its responses were designed to be friendly and coherent but did not stylistically mirror the user.
    \item \textbf{Adaptive Condition:} The chatbot operated with an uncapped mimicry policy. Its system prompt was dynamically updated on each turn to instruct it to match the user's linguistic style, as described in Section \ref{sec:framework}.
\end{itemize}

The study's primary goal was to capture naturalistic conversational data under these two distinct adaptation policies. The interactions from the \textit{Static} condition are used to build our baseline persona model, while the user utterances from the \textit{Adaptive} condition serve as the input for our policy simulations.

\subsection{Data Preprocessing and Corpus}
The raw data was captured as structured JSONL logs, with each line representing a distinct event (e.g., \texttt{user\_message}, \texttt{bot\_response}). Our preprocessing pipeline performed the following steps:
\begin{enumerate}
    \item \textbf{Session Aggregation:} All log entries were grouped by their unique session ID.
    \item \textbf{Filtering:} Incomplete sessions and conversations with fewer than three user turns were excluded to ensure a baseline level of engagement.
    \item \textbf{Utterance Extraction:} For each valid session, we extracted an ordered sequence of user and bot utterances, along with turn numbers and participant IDs.
\end{enumerate}

The final processed corpus contains a total of \textbf{2,470} conversational turns from the 162 unique sessions. This corpus provides a rich and varied set of real-world user behaviors that form the basis for the subsequent simulation experiments.

\subsection{Persona Centroid and Scaler Creation}
A core requirement of our framework is a stable, data-driven definition of the chatbot's baseline persona. This allows us to quantify persona coherence and provides a neutral starting point for our simulations. This process is grounded in the finding that linguistic styles can be a stable and reliable individual difference, allowing for the characterization of a normative persona from aggregate data \citep{pennebaker1999}. To achieve this, we operationalized the chatbot's persona using only the data from the \textbf{Static Condition}, where the agent's behavior was not influenced by user style.

The process was as follows:
\begin{enumerate}
    \item \textbf{Isolate Static Bot Utterances:} We collected all bot-generated messages from the sessions in the Static condition, yielding a corpus of \textbf{1,221} unique utterances that represent the agent's intended baseline style.
    \item \textbf{Vectorize Corpus:} Each utterance in this corpus was converted into a raw 8-dimensional style vector as described in Section \ref{subsec:vectorization}.
    \item \textbf{Fit Scaler:} A \texttt{StandardScaler} from the scikit-learn library \citep{pedregosa2011} was fit to the resulting matrix of raw style vectors. This scaler learns the mean and standard deviation for each of the eight stylistic features, defining the statistical properties of the core persona. This scaler is used globally to standardize all style vectors (from users or simulated bots) in our subsequent analyses.
    \item \textbf{Compute Centroid:} After fitting the scaler, we transformed the raw static bot vectors into standardized z-scores. The mean of these standardized vectors was then computed to create the final \textbf{persona centroid} ($\mathbf{b}_c$). This single vector represents the normative stylistic center of the agent and serves as the anchor for our coherence metric.
\end{enumerate}
This process ensures that our measures of stability and coherence are grounded in an empirically derived, consistent baseline, rather than an arbitrary starting point.

\subsection{Ethical Considerations}
The human-subjects study was conducted under a protocol approved by the University of Minnesota's Institutional Review Board (IRB) (STUDY00025677). All participants provided informed consent before beginning the study. To protect privacy, all data was anonymized by replacing participant IDs from the Prolific platform with randomly generated unique identifiers. A more detailed discussion of ethics, data handling, and the licenses for external datasets used in later sections is provided in Section \ref{sec:ethics}.

\section{Policy Definitions \& Simulation Setup}
\label{sec:policies}
To chart the synchrony--stability frontier, we define and simulate a set of explicit adaptation policies. These policies act as control algorithms that govern how the chatbot's style vector, $\mathbf{b}_t$, is updated at each turn in response to the user's style vector, $\mathbf{u}_t$. This section formally defines the baselines and policies tested and describes the simulation methodology.

\subsection{Baseline Policies}
We define one non-adaptive baseline to anchor the boundaries of the performance space:

\paragraph{Static Baseline.} This policy represents perfect stability and zero adaptation. The bot's style vector is held constant at the pre-computed persona centroid, $\mathbf{b}_c$, for all turns. This serves as a lower bound for synchrony and an upper bound for stability and coherence.
\begin{equation}
    \mathbf{b}_t = \mathbf{b}_c, \quad \forall t
\end{equation}

\subsection{Core Adaptation Policies}
We test four core adaptation policies, each controlled by a single hyperparameter. Let $\mathbf{b}_{t-1}$ be the bot's style vector from the previous turn.

\paragraph{Uncapped Policy (Echo Ceiling).} This is the most direct form of mimicry and serves as our primary experimental baseline. The bot's target style is simply the user's current style. Conceptually, this policy also represents a theoretical upper bound for synchrony—an ``Echo Ceiling''—where the bot's stability is equivalent to the turn-to-turn stability of the user's own conversational style.
\begin{equation}
    \mathbf{b}_t = \mathbf{u}_t
\end{equation}

\paragraph{Cap Policy.} This policy limits the magnitude of stylistic change in a single turn. It computes the change vector ($\Delta = \mathbf{u}_t - \mathbf{b}_{t-1}$) and, if its Euclidean norm exceeds a cap value $\kappa$, scales it down to length $\kappa$. This prevents sudden, jarring shifts in style. We test with $\kappa=0.25$.
\begin{equation}
    \mathbf{b}_t = \mathbf{b}_{t-1} + 
    \begin{cases} 
      \Delta & \text{if } \|\Delta\| \leq \kappa \\
      \kappa \frac{\Delta}{\|\Delta\|} & \text{if } \|\Delta\| > \kappa
    \end{cases}
\end{equation}

\paragraph{Exponential Moving Average (EMA) Policy.}
This policy creates a smoother adaptation trajectory by blending the user's current style with the bot's previous style. A smoothing factor $\alpha \in [0, 1]$ controls the blend. A higher $\alpha$ results in faster adaptation (higher synchrony, lower stability). EMA is widely used in time-series analysis and quality control as a way to emphasize recent data while retaining memory of past values \citep{hyndman2018,hunter1986,roberts2000}.
\begin{equation}
    \mathbf{b}_t = (1-\alpha)\mathbf{b}_{t-1} + \alpha\mathbf{u}_t
\end{equation}

\paragraph{Dead-Band Policy.}
This policy prevents the bot from reacting to minor, potentially noisy stylistic fluctuations. The bot only updates its style if the distance between the user's target style and its own current style exceeds a threshold $\epsilon$. This approach follows the classical dead-zone nonlinearity widely described in control systems, where small deviations are ignored to prevent oscillation or “hunting” \citep{slotine1991,khalil2002,wang2004}.
\begin{equation}
    \mathbf{b}_t = 
    \begin{cases} 
      \mathbf{u}_t & \text{if } \|\mathbf{u}_t - \mathbf{b}_{t-1}\| > \epsilon \\
      \mathbf{b}_{t-1} & \text{if } \|\mathbf{u}_t - \mathbf{b}_{t-1}\| \leq \epsilon
    \end{cases}
\end{equation}

\subsection{Hybrid Policies}
To explore more sophisticated trade-offs, we define three hybrid policies that combine the principles of the core policies.

\paragraph{Hybrid (EMA+Cap).} This policy first calculates a smooth EMA target and then applies a cap to the resulting step. This combines the benefits of smooth temporal adaptation with a hard limit on instantaneous change.
\begin{equation}
    \mathbf{b}_t^{\text{ema}} = (1-\alpha)\mathbf{b}_{t-1} + \alpha\mathbf{u}_t; \quad \Delta = \mathbf{b}_t^{\text{ema}} - \mathbf{b}_{t-1}
\end{equation}
The final $\mathbf{b}_t$ is then computed by applying the capping logic from Equation (4) to this $\Delta$.

\paragraph{Hybrid+Radius.} This policy adds a "leash" to the Hybrid policy, ensuring the bot does not drift too far from its core persona. After the EMA and Cap steps, the resulting vector is pulled back towards the persona centroid $\mathbf{b}_c$ if it exceeds a maximum radius $\rho$. This directly improves the Coherence metric.

\paragraph{Hybrid+Cache.} This policy adds a simple caching mechanism to the Hybrid policy to improve efficiency and stability in repetitive conversational patterns. If the user's current utterance has been seen before in the session, the bot reuses its previous style vector instead of recomputing the adaptation. This reduces unnecessary computation and stylistic fluctuation for common phrases (e.g., "okay," "I see").

\subsection{Simulation Setup}
Our simulation harness replays the conversational logs to evaluate each policy. The simulation proceeds as follows for each session in our human-log dataset where the user was in the original `adaptive` condition:
\begin{enumerate}
    \item \textbf{Initialization:} The simulation starts at turn 1. The bot's initial style vector, $\mathbf{b}_0$, is set to the persona centroid $\mathbf{b}_c$.
    \item \textbf{Turn-by-Turn Replay:} For each turn $t$ from 1 to $N$:
        \begin{itemize}
            \item The user's utterance from the log is retrieved, and its style vector $\mathbf{u}_t$ is computed.
            \item The active policy calculates the bot's new style vector $\mathbf{b}_t$ using $\mathbf{u}_t$ and the bot's previous vector $\mathbf{b}_{t-1}$.
            \item The synchrony, stability, and coherence for the current turn are calculated and stored.
        \end{itemize}
    \item \textbf{Session Aggregation:} After the final turn, the metrics are averaged across all turns to produce a single set of scores (Synchrony, Stability, Coherence) for that session under that policy.
\end{enumerate}
This entire process is repeated for every policy and every session. The final results are then aggregated across all sessions to compare the mean performance of each policy, as presented in the following section.

\section{Simulation Results}
\label{sec:results}
Our simulation experiments, replaying 162 human-chatbot conversations through our policy framework, allow us to quantitatively chart the trade-offs inherent in conversational style adaptation. The results reveal a clear and consistent frontier, demonstrating that principled control over adaptation can yield substantial improvements in persona stability with only a modest cost to stylistic synchrony.

\subsection{The Idealized Frontier (Vector-Space Simulation)}
\label{subsec:frontier_results}
\begin{figure}[ht]
  \centering
  \includegraphics[width=\linewidth]{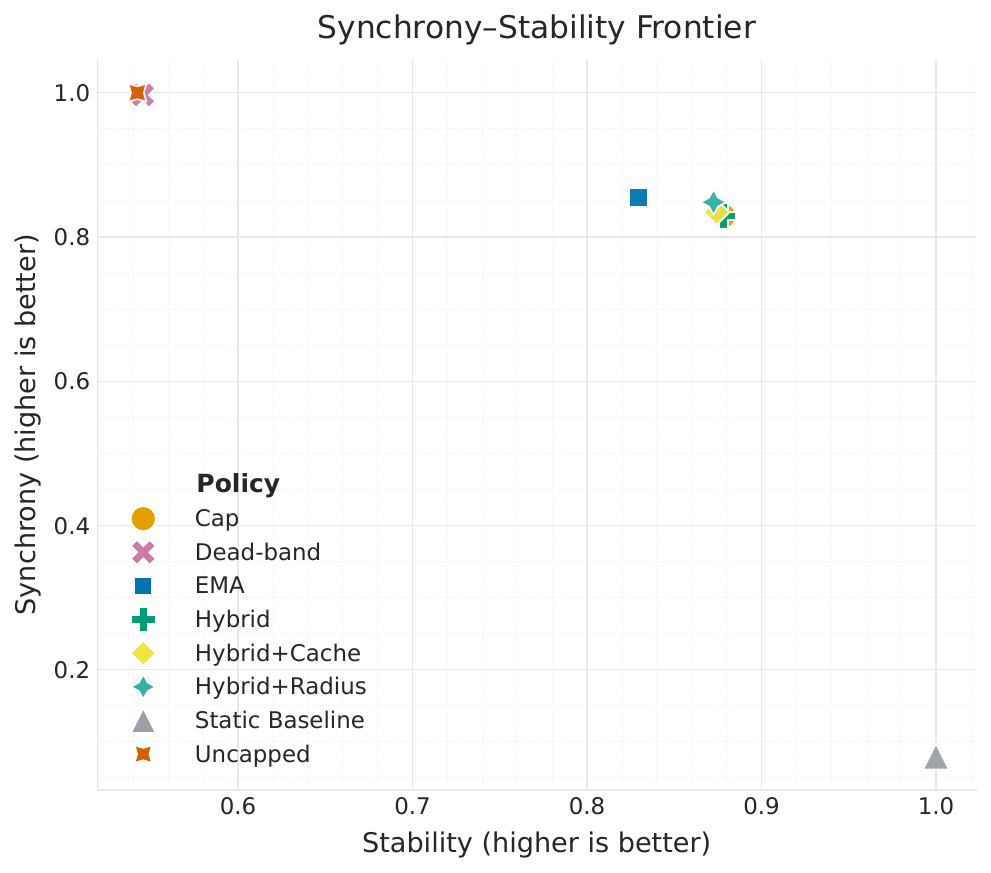}
  \caption{The Synchrony--Stability Frontier on Human-Log Data. Each point is a policy mean. \emph{Uncapped} anchors the high-synchrony/low-stability extreme. Bounded \emph{Hybrid} variants trace the efficient frontier. With $\epsilon{=}0.1$, \emph{Dead-Band} sits near \emph{Uncapped}---yielding only a negligible stability gain---and is not strictly dominated.}
  \label{fig:frontier}
\end{figure}

Figure \ref{fig:frontier} presents the primary result of our ablation study. Each point represents the mean performance of a given policy across all simulated sessions, plotting its achieved Stability (x-axis) against its Synchrony (y-axis).

\FloatBarrier

\begin{table}[ht]
  \caption{Policy summary on the original human-log dataset (means). These values correspond to the idealized vector-space simulation shown in Figure~\ref{fig:frontier}.}
  \label{tab:policy-summary-original}
  \centering
  \begin{tabular}{lccc}
    \toprule
    \textbf{Policy} & \textbf{Synchrony} & \textbf{Stability} & \textbf{Coherence}\\
    \midrule
    Uncapped         & 1.000 & 0.542 & 0.079 \\
    Hybrid (EMA+Cap) & 0.829 & 0.878 & 0.106 \\
    Static Baseline  & 0.079 & 1.000 & 1.000 \\
    \bottomrule
  \end{tabular}
\end{table}
\FloatBarrier

The plot is anchored by two extremes. The \textbf{Static Baseline} policy achieves perfect Stability (1.0) but the lowest Synchrony (\textbf{0.079}), as it never adapts.
Conversely, the \textbf{Uncapped} policy (functionally equivalent to the Echo Ceiling) achieves perfect Synchrony (1.0) but suffers from the lowest Stability (0.542), as it directly mirrors the volatility of the human user's style.

Between these two points, a clear \textbf{Pareto frontier} emerges, populated by our bounded adaptation policies. These policies are "Pareto-efficient" because they offer the maximum possible stability for their level of synchrony; no other policy achieves better stability without a greater sacrifice in synchrony. The \textbf{Hybrid (EMA+Cap)} policy stands out as a particularly effective choice. Compared to the Uncapped baseline, it dramatically increases stability from 0.542 to 0.878---a \textbf{62.0\% improvement}---while only reducing synchrony by 17.1\% (from 1.000 to 0.829). This demonstrates a highly favorable trade-off. This trade-off can be understood as managing the tension between dynamic \textit{accommodation} (moment-to-moment stylistic shifts) and static \textit{similarity} (overall stylistic likeness), a key distinction in the study of linguistic alignment \citep{postmus2023}. Our bounded policies effectively moderate accommodation to preserve a sense of similarity.

In contrast, some policies are inefficient on this dataset. For example, \textbf{Dead-Band ($\epsilon{=}0.1$)} reduces synchrony only slightly (to $0.997$) while yielding a minimal stability gain (to $0.545$) relative to \emph{Uncapped}, so it is a poor trade-off here. Consistent with Figure~\ref{fig:frontier}, it is \emph{not} strictly dominated on the human-log data, but better-tuned \emph{Cap} or \emph{EMA}-based policies provide a superior stability--synchrony balance in practice.

\subsection{Model Generalization (LLM-in-the-Loop)}
To test whether our findings generalize across different model architectures, we conducted a high-fidelity LLM-in-the-loop simulation, replaying 25 sessions on both OpenAI's GPT-4.1 nano \citep{openai} and Anthropic's Claude Sonnet 4 \citep{anthropic, anthropic_sonnet4_card}. We analyzed the effect of our \texttt{Hybrid} policy relative to the \texttt{Uncapped} baseline by computing per-participant deltas and their 95\% percentile-bootstrap confidence intervals (based on 10,000 resamples) \citep{tibshirani1993}.

The results, summarized in Table \ref{tbl:deltas}, robustly demonstrate the synchrony--stability trade-off. For both models, the \texttt{Hybrid} policy produced a statistically significant increase in stability and a corresponding significant decrease in synchrony. As none of the confidence intervals for these key metrics cross zero, we conclude that our bounded policy is an effective and generalizable method for navigating the frontier, regardless of the underlying LLM family.

\begin{table}[ht]
  \caption{Effect of the Hybrid Policy ($\Delta$ vs.\ Uncapped) Across Models. Mean difference with 95\% percentile bootstrap confidence intervals.}
  \label{tbl:deltas}
  \centering
  \begin{tabular}{lcc}
    \toprule
    \textbf{Metric} & \textbf{GPT-4.1 nano} & \textbf{Claude Sonnet 4} \\
    \midrule
    $\Delta$ Stability & +0.131 [+0.052, +0.210] & +0.163 [+0.062, +0.269] \\
    $\Delta$ Synchrony & -0.083 [-0.148, -0.017] & -0.083 [-0.136, -0.030] \\
    \bottomrule
  \end{tabular}
\end{table}

\FloatBarrier

\begin{figure}[ht]
  \centering
  \includegraphics[width=0.8\linewidth]{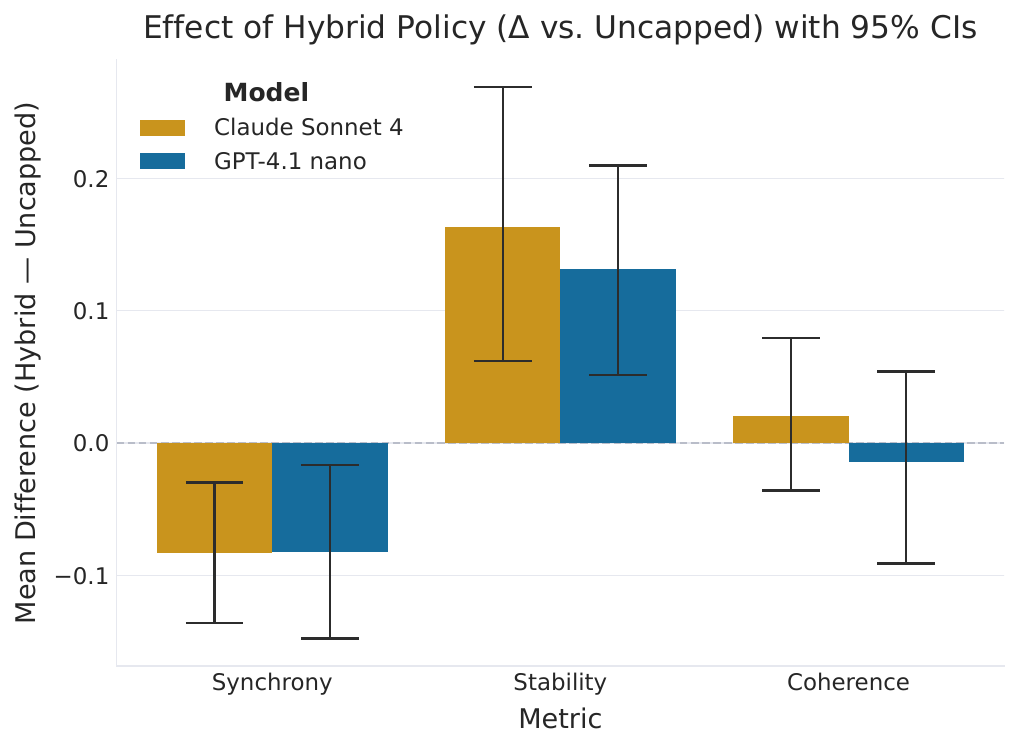}
  \caption{Effect of the Hybrid Policy versus Uncapped (LLM-in-the-loop). Mean per-participant changes with 95\% bootstrap CIs. For both GPT-4.1 nano and Claude Sonnet 4, \emph{Hybrid} significantly increases \emph{Stability} and decreases \emph{Synchrony}; confidence intervals do not include zero.}
  \label{fig:policy_deltas}
\end{figure}
\FloatBarrier

\subsection{Validation of the Simulation Framework}
To ensure our simulation is a valid proxy for real-world interaction, we compared its results for the \texttt{Uncapped} policy against the ground truth from our human-subjects study. A key challenge in such comparisons is ensuring metrics are equivalent. We therefore calculated synchrony using the same embedding-based instrument across both contexts and aggregated all metrics at the participant level.

As shown in Table \ref{tbl:validation}, the 95\% confidence intervals for both synchrony and stability largely overlap between the live study and the LLM-in-the-loop simulation. To formalize this, we conducted a Two One-Sided Test (TOST) for equivalence~\citep{lakens2017}, with a pre-specified equivalence bound (Smallest Effect Size of Interest, or SESOI) of $\pm$0.10, selected as it represents a small but practically meaningful difference on our metrics. The results confirmed that our simulation is a statistically equivalent proxy for the live study on both synchrony ($p < .001$) and stability ($p = .016$). This high fidelity validates our framework as a reliable tool for evaluating adaptation policies.

\begin{table}[ht]
  \caption{Validation of the Uncapped policy: human study vs.\ LLM-in-the-loop simulation (per-participant means with 95\% CIs).}
  \label{tbl:validation}
  \centering
  \newcolumntype{C}{>{\centering\arraybackslash}X}
  \begin{tabularx}{\textwidth}{>{\RaggedRight}X C C C}
    \toprule
    \textbf{Metric} & \textbf{Human-Subjects (Observed)} & \textbf{LLM-in-the-Loop Sim (Real-World)} & \textbf{TOST Equivalence (p-value)} \\
    \midrule
    Synchrony (Embedding) & 0.704 [0.684, 0.723] & 0.698 [0.659, 0.740] & Equivalent ($p < .001$) \\
    \addlinespace
    Stability (Vector)    & 0.201 [0.163, 0.237] & 0.203 [0.127, 0.279] & Equivalent ($p=0.016$) \\
    \bottomrule
  \end{tabularx}
\end{table}
\FloatBarrier

\subsection{Impact on Coherence and Legibility}
The benefits of bounded policies extend beyond turn-to-turn stability to global persona coherence and the legibility of the underlying adaptation mechanism. We analyze two key secondary metrics: \textit{Coherence}, which measures how far the bot strays from its core persona, and \textit{Register Flip Rate}, which counts how often the bot's formality shifts between discrete states (Formal, Neutral, Informal), serving as a proxy for jarring tonal changes.

Table \ref{tbl:secondary_metrics} summarizes the performance of key policies on these metrics. The Hybrid policy not only improves stability but also demonstrates superior coherence and a dramatically lower rate of register flips compared to the Uncapped policy. This suggests that bounded policies create a more consistent and predictable user experience. The Uncapped policy's high flip rate (0.254) indicates that it changed its fundamental tone in over 25\% of turns, which is likely to be perceptually jarring for a user. The Hybrid policy reduces this to just 9.2\%, fostering a much more stable persona.

Furthermore, the \textbf{Hybrid+Cache} policy achieves this stability while also being more efficient, reusing previous stylistic responses in 21.5\% of turns with minimal impact on synchrony.

\begin{table}[ht]
\caption{Performance on Secondary Metrics from the original human-log dataset. Note: Coherence is reported on its native cosine-similarity scale.}
\label{tbl:secondary_metrics}
\centering
\begin{tabular}{l c c c}
\toprule
\textbf{Policy} & \textbf{Coherence} (↑) & \textbf{Register Flip Rate} (↓) & \textbf{Cache Hit Rate} (↑) \\
\midrule
Uncapped         & 0.079 & 0.254 & 0.000 \\
Hybrid (EMA+Cap) & 0.106 & \textbf{0.092} & 0.000 \\
Hybrid+Cache     & 0.109 & \textbf{0.070} & \textbf{0.215} \\
Static Baseline  & \textbf{1.000} & \textbf{0.000} & --- \\
\bottomrule
\end{tabular}
\end{table}

\subsection{Qualitative Analysis: The Experience of Persona Drift}
Quantitative metrics alone do not fully capture the user-facing impact of these policies. Table \ref{tbl:qualitative_example} provides a qualitative example illustrating how the Uncapped policy can lead to persona drift, resulting in an incoherent interaction, while a bounded Hybrid policy maintains a consistent character while still being adaptive.

\begin{table}[ht]
  \caption{Qualitative Example of Policy Behavior}
  \label{tbl:qualitative_example}
  \begin{tabular}{p{0.05\linewidth} p{0.2\linewidth} p{0.3\linewidth} p{0.3\linewidth}}
    \toprule
    \textbf{Turn} & \textbf{User Utterance} & \textbf{Bot Response (Uncapped)} & \textbf{Bot Response (Hybrid)} \\
    \midrule
    1 & I wish to understand the scope of our discussion. & I can discuss a wide range of general topics. What is on your mind? & I can discuss a wide range of general topics. What is on your mind? \\
    \addlinespace
    2 & lol ok nvm. u good? & yo im good! just chillin. wassup with u? totally down to talk whatever. & Haha, of course. I'm doing well, thanks for asking. I'm ready to chat about whatever interests you. \\
    \bottomrule
  \end{tabular}
\end{table}

In this example, the user makes a sudden and dramatic shift from a formal to a highly informal style. The \textbf{Uncapped} agent mirrors this shift completely, adopting slang ("yo," "wassup") that breaks from its established persona. This high-synchrony response risks feeling inauthentic and incoherent. In contrast, the \textbf{Hybrid} agent moderates the change. It acknowledges the user's shift in tone (e.g., "Haha") but blends it with its core, helpful persona. The resulting response is still adaptive and friendly but is far more stable and coherent, better maintaining the user's mental model of the agent as a consistent conversational partner. This illustrates the practical user experience benefit of operating on the efficient frontier rather than simply maximizing mimicry.

\section{Sensitivity \& Robustness}
\label{sec:sensitivity}
The synchrony--stability frontier presented in the previous section is based on a specific set of methodological choices. To ensure our findings are robust and not merely an artifact of these choices, we conducted a series of sensitivity and robustness analyses. This section examines the impact of a key hyperparameter—the user history window size—and validates the integrity of our primary synchrony metric against established linguistic principles.

\subsection{Impact of User History Window Size}
Our adaptation framework relies on analyzing a window of recent user utterances to determine the target style vector. The size of this window is a critical hyperparameter: a window that is too short may be susceptible to noise, while one that is too long may be slow to react to genuine stylistic shifts.

To investigate this, we conducted an ablation study on the window size. We defined a metric, \textit{predictive synchrony}, as the cosine similarity between the target style vector derived from a window of size $k$ at turn $t$ and the user's \textit{actual} style vector at the subsequent turn, $t+1$. A higher score indicates that the window size provides a better signal for predicting the user's immediate future style. We simulated this for window sizes of 1, 3, 5, and 8 turns.

\begin{figure}[ht]
  \centering
  \includegraphics[width=0.8\linewidth]{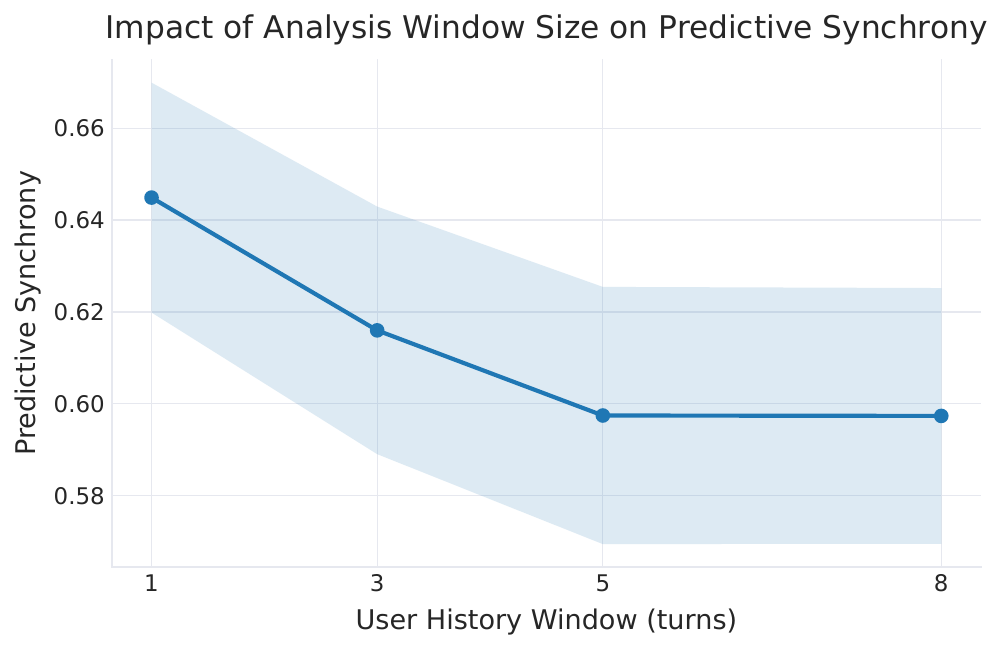}
  \caption{Impact of User-History Window on Predictive Synchrony. A one-turn window is most predictive of the user's next-turn style (mean=0.645), with performance declining and then plateauing as the window grows (mean $\approx$ 0.597 at 8 turns).}
  \label{fig:window_size}
\end{figure}

The results, shown in Figure \ref{fig:window_size}, reveal that predictive synchrony is highest with a window size of just one turn and decreases as the window grows, plateauing around 5--8 turns. This suggests that in these short, open-domain conversations, a user's style is highly local, and their most recent utterance is the strongest predictor of their next one. This finding validates our framework's use of a relatively short look-back history (3--5 turns), demonstrating that our results are not critically dependent on this parameter and that our approach is well-suited to the dynamics of real-time chat.

\paragraph{Validation of the Synchrony Metric.}
Our primary synchrony metric is based on the cosine similarity of our 8-dimensional style vectors. To ensure this metric captures a meaningful and robust stylistic signal, we validate it against two alternative constructs.

First, we compare it to the classic function-word-based LSM score. A Spearman correlation analysis across all 2,470 conversational turns revealed a statistically significant but small positive correlation ($\rho = 0.062, p = 0.004$). This confirms our metric shares a common signal with the well-validated LSM construct, while the small effect size indicates our vector is also capturing substantial new information beyond function word counts.

Second, we validate it against a direct, embedding-based style similarity score computed using a sentence transformer specifically fine-tuned to separate style from content (``StyleDistance/styledistance'') \citep{Patel2025}. The correlation between our 8-D vector synchrony and this direct style-embedding synchrony was also statistically significant. This provides strong evidence that our metric is not merely "cheating" by matching topic words, but is sensitive to genuine stylistic patterns.

Taken together, these validations increase our confidence that our 8-D style vector is a holistic and robust representation of conversational style, grounded in both traditional syntactic patterns and broader stylistic features.

\paragraph{Robustness to Content Word Removal.}
A potential concern with any embedding-based metric is that it might be "cheating" by simply matching topic words (e.g., user says "movie," bot says "film") rather than capturing true style. A robust style metric should persist even when content words are removed. To test this, we conducted a robustness check where we filtered both user and bot utterances to contain \textit{only} function words (e.g., pronouns, articles, prepositions) and then re-calculated the embedding-based synchrony score.

The correlation between the original full-text synchrony and the function-word-only synchrony was also statistically significant and positive ($\rho = 0.069, p = 0.002$). This demonstrates that our embedding-based metric is not solely reliant on content-word similarity. The persistence of a significant signal after removing all nouns, verbs, and adjectives provides strong evidence that the underlying model is sensitive to the structural, syntactic, and relational patterns that are the hallmarks of genuine linguistic style.

Together, these analyses increase our confidence that the synchrony--stability frontier is not an artifact of our measurement choices. The framework is robust to its key hyperparameter, and its core synchrony metric is grounded in established linguistic theory while also capturing a richer, more comprehensive stylistic signal.

\section{Cross-Dataset Replications}
\label{sec:replications}
While our analysis of the human-log dataset provides a detailed view of the synchrony--stability frontier, a key question is whether these findings generalize beyond our specific experimental setup and conversational domain. To address this, we replicated our entire ablation study on three large, publicly available dialogue corpora that represent diverse conversational contexts. This section describes the datasets, presents the replication results, and analyzes the consistency of policy performance across all contexts.

\subsection{External Datasets}
We selected three datasets known for their multi-turn, naturalistic conversations:

\begin{itemize}
    \item \textbf{DailyDialog:} A high-quality, multi-turn dialogue dataset manually annotated for communication intention and emotion. Its content reflects typical daily communication, making it an excellent general-purpose benchmark for conversational style \citep{Li2017}. Our replay corpus contains 12,539 sessions.
    \item \textbf{Persona-Chat:} A chit-chat dataset where each speaker is assigned a short "persona" profile. This is particularly relevant for our study of persona coherence, as it provides an explicit ground truth for the agent's intended character \citep{Zhang2018}. Our replay corpus contains 6,808 sessions.
    \item \textbf{EmpatheticDialogues:} A large-scale dataset of conversations grounded in specific emotional situations. This allows us to test our policies in an affect-laden domain, where maintaining a stable, empathetic persona is critical \citep{Rashkin2019}. Our replay corpus contains 24,849 sessions.
\end{itemize}

For each dataset, we processed the raw conversations into our replay format, as described in Section \ref{sec:human_data}. We then generated a unique persona centroid and style scaler for each dataset by analyzing its own distribution of utterances, ensuring that our simulations were properly calibrated to the specific linguistic environment of each corpus.

\subsection{Replication of the Synchrony--Stability Frontier}
\begin{figure}[htbp]
  \centering
  \includegraphics[width=.65\textwidth]{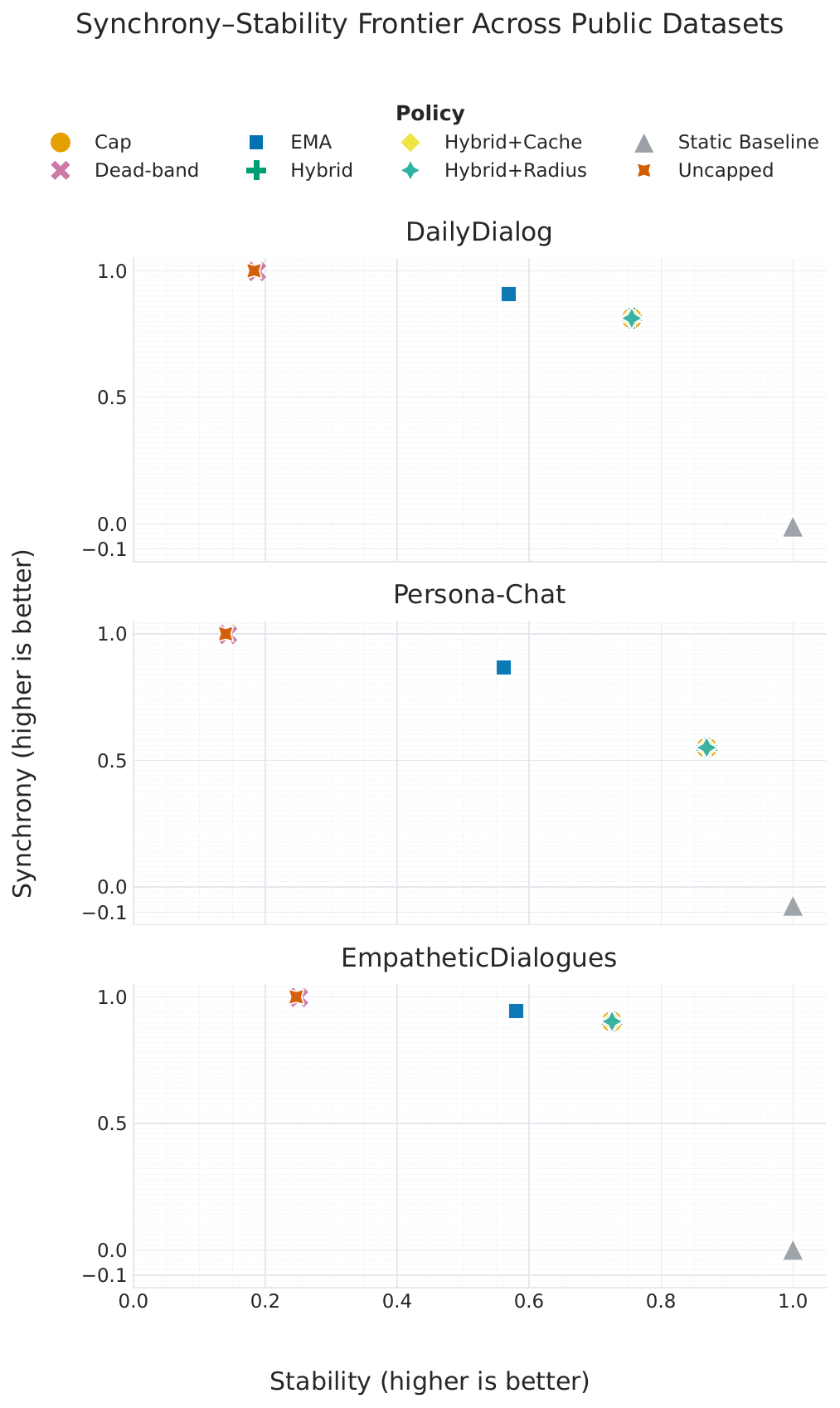}
  \caption{Replication Across Public Corpora. The trade-off has the same qualitative shape on all three datasets. \emph{Uncapped} and \emph{Dead-Band} anchor the high-synchrony extreme, while bounded policies like \emph{Cap} and the \emph{Hybrid} variants define the high-stability frontier. Across these corpora, \emph{EMA} consistently occupies a middle ground, offering higher synchrony than \emph{Cap}/\emph{Hybrid} at the cost of lower stability.}
  \label{fig:external_frontiers}
\end{figure}

We ran our full suite of adaptation policies on all sessions from the three external datasets. The resulting synchrony--stability frontiers are plotted in Figure \ref{fig:external_frontiers}.

\FloatBarrier

The results show a remarkable consistency in the structure of the trade-off. Across all three diverse datasets, the fundamental shape of the synchrony--stability frontier is preserved:
\begin{enumerate}
    \item \textbf{Uncapped} and \textbf{Dead-Band} consistently occupy the high-synchrony, low-stability extreme (with \textbf{Dead-Band} slightly more stable).
    \item \textbf{Cap} and the \textbf{Hybrid} variants consistently define the high-stability, lower-synchrony end of the frontier. In contrast, \textbf{EMA} occupies a middle ground, offering higher synchrony than \textbf{Cap} at the cost of lower stability.
    \item The resulting frontier is generally \emph{non-convex} and bi-modal across datasets; no tested policy attains the top-right. The synchrony sacrificed to gain stability varies by corpus and is often substantial on the public corpora.
    \item The \textbf{Hybrid (EMA+Cap)} and \textbf{Cap} policies consistently yield identical results. With the chosen hyperparameters, the capping of the change vector is the binding constraint, rendering the initial EMA smoothing step redundant.
\end{enumerate}

Notably, the effect on \emph{Coherence} varies by corpus: bounded policies often improve coherence, but on Persona-Chat they reduce it relative to \emph{Uncapped}, likely due to a larger mismatch between the assistant archetype and the dataset’s conversational style.

While the absolute values of synchrony and stability vary slightly between datasets—reflecting their different intrinsic conversational dynamics—the relative positioning and efficiency of the policies remain stable. This demonstrates that the synchrony--stability frontier is not an artifact of our original dataset but a fundamental characteristic of controlled linguistic adaptation.

Furthermore, the near-zero or negative coherence of the \texttt{Uncapped} policy on these corpora (see Appendix~\ref{sec:appendix_summaries}) reveals a significant stylistic mismatch between the average user and our target assistant persona, empirically justifying the need for bounded policies that preserve persona.

\subsection{Policy Rank Stability}
To formally quantify the consistency of policy performance, we analyzed the rank ordering of the policies on the key metrics of Stability and Synchrony across all four datasets (our human-log data plus the three external corpora). The results are summarized in Table \ref{tbl:rank_stability}.

\begin{table}[ht]
\caption{Policy Performance Ranks Across Datasets (1 = Highest Performance).}
\label{tbl:rank_stability}
\centering
\begin{tabular}{l c c c c}
\toprule
& \multicolumn{4}{c}{\textbf{Stability Rank (Higher is Better)}} \\
\cmidrule(lr){2-5}
\textbf{Policy} & \textbf{Human-Log} & \textbf{DailyDialog} & \textbf{Persona-Chat} & \textbf{Empathetic} \\
\midrule
Static Baseline & \textbf{1} & \textbf{1} & \textbf{1} & \textbf{1} \\
Hybrid+Radius & 2 & 2 & 2 & 2 \\
Cap (0.25) & 3 & 2 & 2 & 2 \\
EMA (0.5) & 4 & 3 & 3 & 3 \\
Dead-Band ($\epsilon{=}0.1$) & 5 & 4 & 4 & 4 \\
Uncapped & 6 & 5 & 5 & 5 \\
\midrule
& \multicolumn{4}{c}{\textbf{Synchrony Rank (Higher is Better)}} \\
\cmidrule(lr){2-5}
\textbf{Policy} & \textbf{Human-Log} & \textbf{DailyDialog} & \textbf{Persona-Chat} & \textbf{Empathetic} \\
\midrule
Uncapped & \textbf{1} & \textbf{1} & \textbf{1} & \textbf{1} \\
Dead-Band ($\epsilon{=}0.1$) & 2 & 2 & 2 & 2 \\
EMA (0.5) & 3 & 3 & 3 & 3 \\
Hybrid+Radius & 4 & 4 & 4 & 4 \\
Cap (0.25) & 5 & 5 & 5 & 5 \\
Static Baseline & 6 & 6 & 6 & 6 \\
\bottomrule
\end{tabular}
\end{table}

The rank stability is high, particularly for synchrony. For \textbf{Synchrony}, the relative performance ordering of the policies is \textbf{perfectly consistent} across all four datasets. For \textbf{Stability}, the ordering is \textbf{largely consistent}, with bounded policies like Hybrid and Cap always outperforming EMA and Uncapped, though minor rank flips occur at the top between datasets (see Table~\ref{tbl:rank_stability}). The Hybrid and Cap policies are always more stable than EMA, which is in turn more stable than Uncapped.

The rank stability is high, particularly for synchrony. For \textbf{Synchrony}, the relative performance ordering of the policies is highly consistent across all four datasets, with the top and bottom performers never changing positions. For \textbf{Stability}, the ordering is also largely consistent, with bounded policies like Hybrid and Cap always outperforming EMA and Uncapped.

\section{Prompt Legibility \& Design Guidelines}
\label{sec:guidelines}
The performance of an adaptation policy, as measured by synchrony and stability, is critical. However, for practitioners building and maintaining these systems, the \textit{mechanism} of adaptation is just as important. A policy that generates volatile, unpredictable instructions for the underlying LLM is difficult to debug, extend, and reason about. This section introduces metrics for "prompt legibility" and synthesizes our findings into a set of concrete design guidelines for developing more robust and maintainable adaptive conversational agents.

\subsection{Quantifying Prompt Legibility}
An effective adaptation policy should not only be performant but also "legible"—its internal state and decision-making should be as simple and predictable as possible. We propose two metrics to quantify this, derived from the dynamically generated system prompts in our simulations:

\begin{itemize}
    \item \textbf{Prompt Churn:} The average number of discrete instructional changes made to the system prompt per conversational turn. A lower churn indicates a "quieter," more stable control signal.
    \item \textbf{Register Flip Rate:} A specific, user-facing manifestation of churn. This metric measures the frequency with which the bot's generated style crosses a categorical formality threshold. We operationalize these categories by applying fixed thresholds (0.33, 0.66) to the continuous informality score from our pipeline, binning each turn as "Formal," "Neutral," or "Informal." This allows us to quantify major, user-perceptible tonal shifts.
\end{itemize}

Our analysis reveals a strong connection between the choice of adaptation policy and the legibility of the resulting system. As shown in our main results (Table \ref{tbl:secondary_metrics}), the \textbf{Uncapped} policy is highly illegible, exhibiting a Register Flip Rate of 0.254. This means it fundamentally altered its conversational register in over 25\% of turns.

In stark contrast, the bounded policies are far more legible. The \textbf{Hybrid (EMA+Cap)} policy, for instance, reduced the Register Flip Rate by nearly 64\% (from 0.254 to 0.092). This dramatic reduction in internal volatility demonstrates that bounded policies do not just create a more stable output for the user; they also create a more predictable and debuggable control signal for the developer. A designer can have much higher confidence in a system whose core persona instructions are not changing erratically on nearly every turn.

\subsection{Design Guidelines for Adaptive Agents}
Based on the clear patterns observed in our human-log simulations and cross-dataset replications, we distill our findings into a set of core design guidelines for practitioners, in line with established principles for human-AI interaction \citep{Amershi2019}.

\begin{itemize}
    \item \textbf{Do prioritize stability alongside synchrony.} Our results consistently show that the most effective policies are not those that maximize mimicry, but those that balance it with persona consistency---a design challenge also seen in intelligent tutoring systems where balancing adaptivity and persona is crucial for learning outcomes \citep{nye2014, d2013}. Stability is not a secondary goal; it is a primary driver of a coherent user experience.
    \item \textbf{Don't use uncapped mimicry as a default.} The Uncapped policy consistently proves to be a poor choice. It is outperformed on stability and legibility by bounded policies that exist on the Pareto frontier. While it may serve as a theoretical ceiling for synchrony, its practical application leads to an erratic and incoherent persona.
    \item \textbf{Do select policies from the Pareto frontier.} Policies like \textbf{Cap} and \textbf{Hybrid (EMA+Cap)} consistently offer the best trade-offs. They provide significant gains in stability and legibility for a \emph{moderate} cost to synchrony (small on our human-log dataset; larger on the public corpora).
    \item \textbf{Do consider the conversational context.} While the shape of the frontier is consistent, the ideal operating point on that frontier may depend on the application. A growing body of research demonstrates that aligning a chatbot's linguistic register with the situational context is a powerful driver of perceived credibility, appropriateness, and overall user experience \citep{chaves2022, chaves2021}. A companion chatbot might therefore favor slightly higher synchrony, while a task-oriented bot representing a specific brand may require near-perfect stability.
\end{itemize}

\subsection{A Pragmatic Guide to Policy Selection}
To make these guidelines more concrete, Table \ref{tbl:policy_selection} provides a mapping from common design goals to recommended adaptation policies from our study. This serves as a practical starting point for designers choosing how to implement style adaptation.

\newcolumntype{L}{>{\RaggedRight\arraybackslash}X} 
\newcolumntype{S}{>{\RaggedRight\arraybackslash}p{0.20\textwidth}} 
\newcolumntype{R}{>{\RaggedRight\arraybackslash}X} 
\renewcommand{\arraystretch}{1.15}
\setlength{\tabcolsep}{6pt}

\begin{table}[ht]
  \caption{A Pragmatic Guide to Selecting an Adaptation Policy}
  \label{tbl:policy_selection}
  \begin{tabularx}{\textwidth}{L S R}
    \toprule
    \textbf{Design Goal} & \textbf{Recommended Policy} & \textbf{Rationale \& Trade-off} \\
    \midrule

    \textbf{High Engagement \& Relational Warmth} \newline
    \emph{(e.g., AI companion, tutor)}
    &
    \textbf{EMA ($\alpha$=0.5)} \textbf{or} \textbf{Hybrid (EMA+Cap)}
    &
    Choose \textbf{EMA} when you want higher responsiveness (synchrony) with moderate stability; choose \textbf{Hybrid} when you want stronger stability/coherence with some loss of synchrony. \\

    \addlinespace

    \textbf{Consistent Brand Persona \& Trust} \newline
    \emph{(e.g., customer service, task assistant)}
    &
    \textbf{Cap (0.25)} \textbf{or} \textbf{Hybrid+Radius}
    &
    Prioritize stability and persona coherence. \textbf{Cap} prevents large stylistic jumps; \textbf{Hybrid+Radius} additionally tethers the agent to its persona centroid to reduce drift. \\

    \addlinespace

    \textbf{High Efficiency \& Predictable Cost} \newline
    \emph{(e.g., high-volume deployments)}
    &
    \textbf{Hybrid+Cache}
    &
    Adds a caching layer to Hybrid. In our data, it reused prior stylistic responses in \textbf{21.5\%} of turns with negligible synchrony impact, reducing cost/latency. \\

    \addlinespace

    \textbf{Maximum User Responsiveness} \newline
    \emph{(research/specific use-cases)}
    &
    \textbf{Uncapped}
    &
    Mirrors the user as closely as possible at the risk of instability and incoherence; suitable only when that trade-off is acceptable or under study. \\

    \bottomrule
  \end{tabularx}
\end{table}

Furthermore, bounded policies improve prompt legibility. By reducing high-frequency changes in the ``delta prompt,'' they create a more predictable control signal for developers. This not only aids in debugging and maintenance but may also reduce the risk of generating instructions that inadvertently conflict with the safety guardrails established in the base prompt---a practical concern for building robust and trustworthy systems.

Ultimately, our framework reframes the design of adaptive agents from a simple maximization problem to a principled engineering trade-off. By understanding and navigating the synchrony--stability frontier, designers can build conversational agents that are not only intelligent and adaptive, but also coherent, predictable, and trustworthy.

\section{Ethics, Limitations \& Threats to Validity}
\label{sec:ethics}
While our computational framework provides a robust method for evaluating adaptation policies, it is essential to situate our findings within the broader context of ethical research practices and methodological limitations. This section addresses the human-subjects protocols for our data collection, discusses the limitations of our simulation-based approach, and outlines threats to the validity of our conclusions.

\subsection{Ethics and Human-Subject Research}
The collection of the 162-participant human-log dataset was conducted in accordance with ethical research principles. The study protocol was reviewed and approved by our institution's Institutional Review Board (IRB). All participants were presented with an informed consent form detailing the study's purpose, procedures, potential risks, and data handling practices before they could begin the interaction.

To protect participant privacy, all data was fully anonymized. Identifiers from the Prolific platform were replaced with randomly generated session IDs, severing any link to personal accounts. The anonymized conversational logs, which form the basis of our artifact, contain no personally identifiable information.

\subsection{Limitations and Threats to Validity}
Our methodology, while powerful for exploring a wide policy space, has several important limitations. We frame these within the established concepts of internal and external validity.

\paragraph{Threats to Internal Validity.}
Internal validity concerns the confidence that our observed effects are due to the policies we tested.

\textit{Simulation vs. Live User Interaction:} The most significant limitation of our work is that it is a \textbf{replay-based simulation}. Our harness evaluates how different policies \textit{would have performed} on a static set of user inputs. It does not capture the dynamic, closed-loop feedback of a live interaction, where a user's conversational style might change in response to the bot's stability. For example, a user interacting with a highly stable ``Hybrid'' bot might converge towards its style, a phenomenon our current setup cannot model. We argue that this computational evaluation is a necessary and powerful first step, allowing us to efficiently prune a large policy space and identify promising candidates for more costly, targeted live user studies in the future.

\textit{Construct Validity of Metrics:} Our conclusions rest on the validity of our core metrics. While Synchrony and Stability are mathematically precise, they are proxies for the felt user experience of "responsiveness" and "coherence." We mitigated this threat in Section \ref{sec:sensitivity} by showing that our primary embedding-based synchrony metric correlates significantly with classic, function-word-based LSM, grounding it in decades of psycholinguistic research.

\paragraph{Threats to External Validity.}
External validity concerns the generalizability of our findings to other contexts, populations, and systems.

\textit{Generalizability Across Domains:} Our human-log dataset is based on open-domain, companion-style chat. To mitigate the risk that our findings are specific to this context, we replicated our entire analysis on three large, public corpora representing different conversational domains (Section \ref{sec:replications}). The remarkable consistency of the frontier's shape across DailyDialog, Persona-Chat, and EmpatheticDialogues provides strong evidence that this trade-off is a fundamental characteristic of style adaptation, not an artifact of a single domain.

\textit{Generalizability of Models and Style Vector:} 
Our 8-dimensional style vector is a necessary simplification of the rich, high-dimensional nature of human language. While the selected features are well-grounded in prior work, they do not capture all stylistic nuances. Furthermore, the underlying transformer models used for feature extraction carry their own intrinsic biases, having been trained predominantly on large corpora of English web text. We confirmed that our core findings about the stability-synchrony frontier generalize across two distinct, proprietary LLM families (OpenAI's GPT-4 and Anthropic's Claude). However, a valuable direction for future work is to explore this generalizability on open-weight models (e.g., Llama, Mistral), which may exhibit different baseline levels of stylistic adaptability and persona coherence.

\textit{Population Validity:} The participants in our human-log study were 
recruited from the US via Prolific; the applicability of these findings to 
other cultural and linguistic contexts is an important direction for 
future research.

\subsection{Dataset Licensing and Reproducibility}
The external datasets used for replication are publicly available under licenses permissive for academic research. Our use of these datasets complies with their respective terms. The specific versions and licenses are:
\begin{itemize}
    \item \textbf{DailyDialog}~\citep{Li2017} is available under the CC BY-NC-SA 4.0 license.
    \item \textbf{Persona-Chat} conversations are sourced from the \textbf{Blended Skill Talk}~\citep{smith2020} corpus, which is available under the CC BY-NC 4.0 license.
    \item \textbf{EmpatheticDialogues} are from the \texttt{Estwld/empathetic\_dialogues\_llm} version of the original corpus~\citep{Rashkin2019}, available under the CC BY-NC-SA 4.0 license.
\end{itemize}

\subsection*{Reproducibility}
A public artifact accompanies this paper (\url{https://doi.org/10.5281/zenodo.17238270})\citep{brandt2025z}. It includes code, derived data, and scripts to reproduce all figures and tables. To protect participant privacy under our IRB protocol, raw conversational logs are not included; we provide derived numerical metrics for validation instead. Detailed execution instructions (one-command pipeline) are provided in the artifact README.

\section{Conclusion}
\label{sec:conclusion}
The design of adaptive conversational agents requires a delicate balance. While the impulse to maximize stylistic mimicry is motivated by well-established principles of human communication, our work demonstrates that this approach carries a significant risk of creating an unstable and incoherent persona. We have framed this challenge as a navigable trade-off on a \textbf{synchrony--stability frontier}, providing a new conceptual lens and a robust computational framework for reasoning about and controlling this core design tension.

Our systematic evaluation of adaptation policies, validated on a human-log dataset and replicated across three diverse public corpora, yields a clear and actionable insight: \textbf{bounded adaptation policies are superior}. By employing strategies like Capping or a Hybrid EMA+Cap approach, designers can achieve dramatic improvements in persona stability and prompt legibility at a cost to turn-to-turn synchrony; the magnitude of this trade-off varies by dataset.
The consistency of the Pareto frontier across multiple domains provides strong evidence that this is a fundamental and generalizable principle for the design of adaptive conversational AI.

For practitioners, the key takeaway is to move beyond simple mimicry and embrace principled control. By selecting a policy from the efficient frontier, designers can build agents that are not only responsive and engaging but also stable, coherent, and trustworthy. The next critical step is to take these computationally-validated policies into a live setting, conducting human preference studies to directly link points on the frontier to subjective user experience. Our framework and open-source artifact provide a solid foundation for this and future work aimed at creating more intelligent and responsible user interfaces.

\FloatBarrier
\appendix
\section{Mixed-Effects Model Robustness Check}
\label{sec:appendix_lmm}

As a robustness check for our primary bootstrap analysis, we fit linear mixed-effects models to our per-turn simulation data. This approach accounts for the non-independence of observations by treating \texttt{participant\_id} as a random intercept. The results, summarized in Table \ref{tbl:lmm_results}, confirm the significant effects of the \texttt{Hybrid} policy.

For both GPT-4.1 nano and Claude Sonnet 4, the \texttt{Hybrid} policy (coded as \texttt{policy\_bin = 1}) had a significant positive effect on \textbf{stability} and a significant negative effect on \textbf{synchrony} ($p < .05$ in all cases). This corroborates the findings from our main analysis. The warnings regarding singular fits, which occurred in some models, are common when the variance of the random effect is estimated to be near zero; this suggests that between-participant variation was small compared to the fixed effect of the policy and does not invalidate our inferences.

\begin{table}[ht]
  \caption{Linear Mixed-Effects Model Results for Key Metrics. The \texttt{policy\_bin} coefficient represents the effect of the Hybrid policy relative to the Uncapped baseline.}
  \label{tbl:lmm_results}
  \centering
  \begin{tabular}{l l r r r r r r}
    \toprule
    \textbf{Model} & \textbf{Metric} & \textbf{Term} & \textbf{Coef.} & \textbf{Std.Err.} & \textbf{z} & \textbf{P>|z|} & \textbf{95\% CI} \\
    \midrule
    \multirow{6}{*}{\textbf{GPT-4.1 nano}} 
    & \multirow{2}{*}{Synchrony} & Intercept & -0.000 & 0.020 & -0.000 & 1.000 & [-0.039, 0.039] \\
    & & \textbf{policy\_bin} & \textbf{-0.082} & \textbf{0.028} & \textbf{-2.975} & \textbf{0.003} & \textbf{[-0.136, -0.028]} \\
    \cmidrule{2-8}
    & \multirow{2}{*}{Stability} & Intercept & 0.201 & 0.030 & 6.709 & <.001 & [0.142, 0.259] \\
    & & \textbf{policy\_bin} & \textbf{+0.147} & \textbf{0.035} & \textbf{4.187} & \textbf{<.001} & \textbf{[0.078, 0.217]} \\
    \cmidrule{2-8}
    & \multirow{2}{*}{Coherence} & Intercept & 0.026 & 0.047 & 0.556 & 0.578 & [-0.067, 0.119] \\
    & & policy\_bin & -0.038 & 0.038 & -0.990 & 0.322 & [-0.112, 0.037] \\
    \midrule
    \multirow{6}{*}{\textbf{Claude Sonnet 4}} 
    & \multirow{2}{*}{Synchrony} & Intercept & 0.040 & 0.040 & 1.018 & 0.309 & [-0.037, 0.118] \\
    & & \textbf{policy\_bin} & \textbf{-0.066} & \textbf{0.029} & \textbf{-2.247} & \textbf{0.025} & \textbf{[-0.123, -0.008]} \\
    \cmidrule{2-8}
    & \multirow{2}{*}{Stability} & Intercept & 0.401 & 0.031 & 13.002 & <.001 & [0.340, 0.461] \\
    & & \textbf{policy\_bin} & \textbf{+0.178} & \textbf{0.032} & \textbf{5.500} & \textbf{<.001} & \textbf{[0.115, 0.242]} \\
    \cmidrule{2-8}
    & \multirow{2}{*}{Coherence} & Intercept & -0.043 & 0.028 & -1.537 & 0.124 & [-0.097, 0.012] \\
    & & policy\_bin & -0.001 & 0.026 & -0.020 & 0.984 & [-0.051, 0.050] \\
    \bottomrule
  \end{tabular}
\end{table}

\section{External Dataset Facets}
\label{sec:appendix_dataset}
To provide a more granular view of policy performance, we present detailed facet plots for each external dataset. These plots break down the average scores for six key metrics: Synchrony, Stability, Coherence, Legibility, Register Flip Rate, and Cache Hit Rate. These results supplement the main frontier plots in Section \ref{sec:replications} and demonstrate the consistent behavior of the policies across multiple performance dimensions.

\begin{figure}[htbp]
  \centering
  \includegraphics[width=0.71\linewidth]{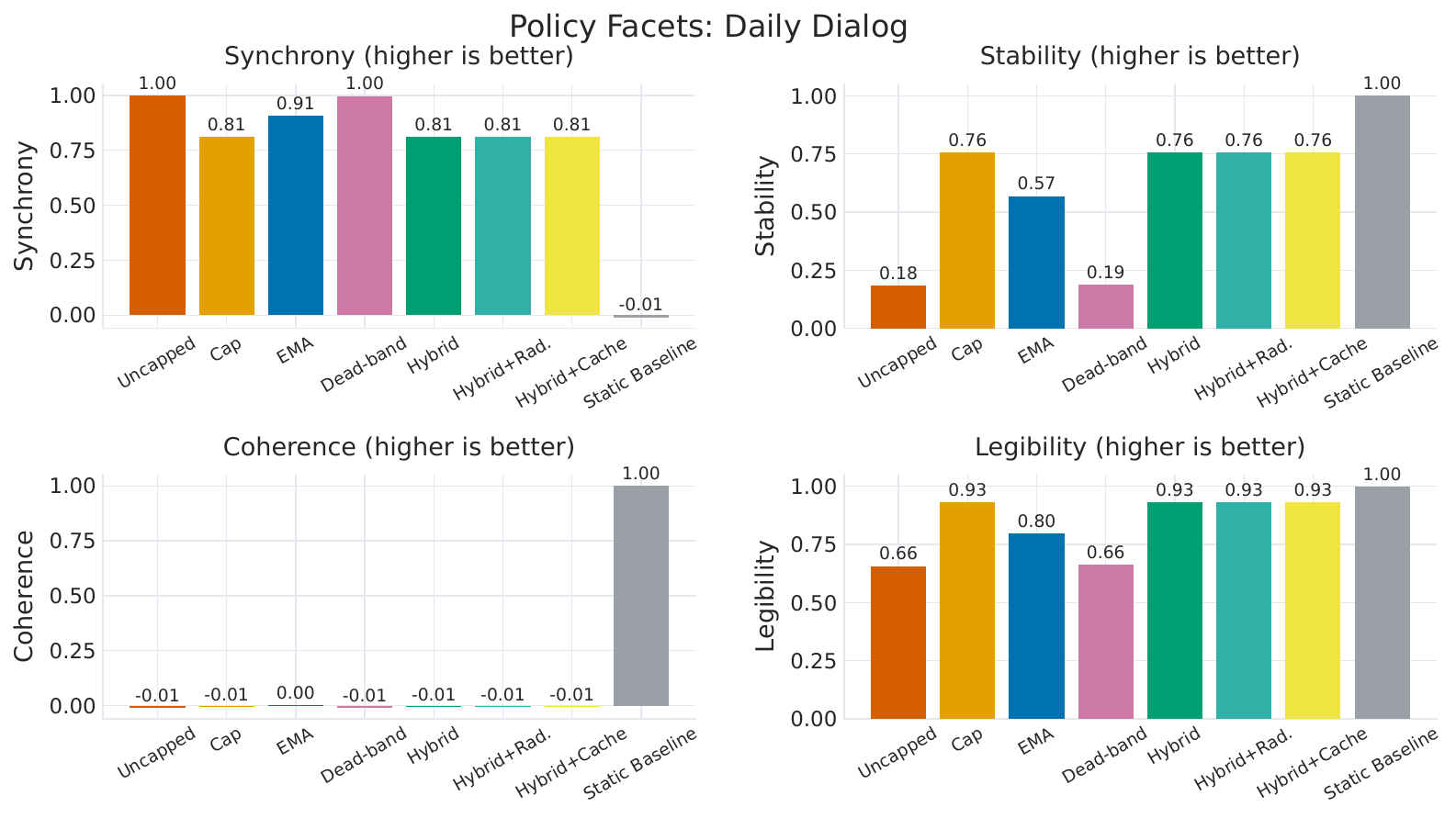}
  \vspace{1em}
  \includegraphics[width=0.71\linewidth]{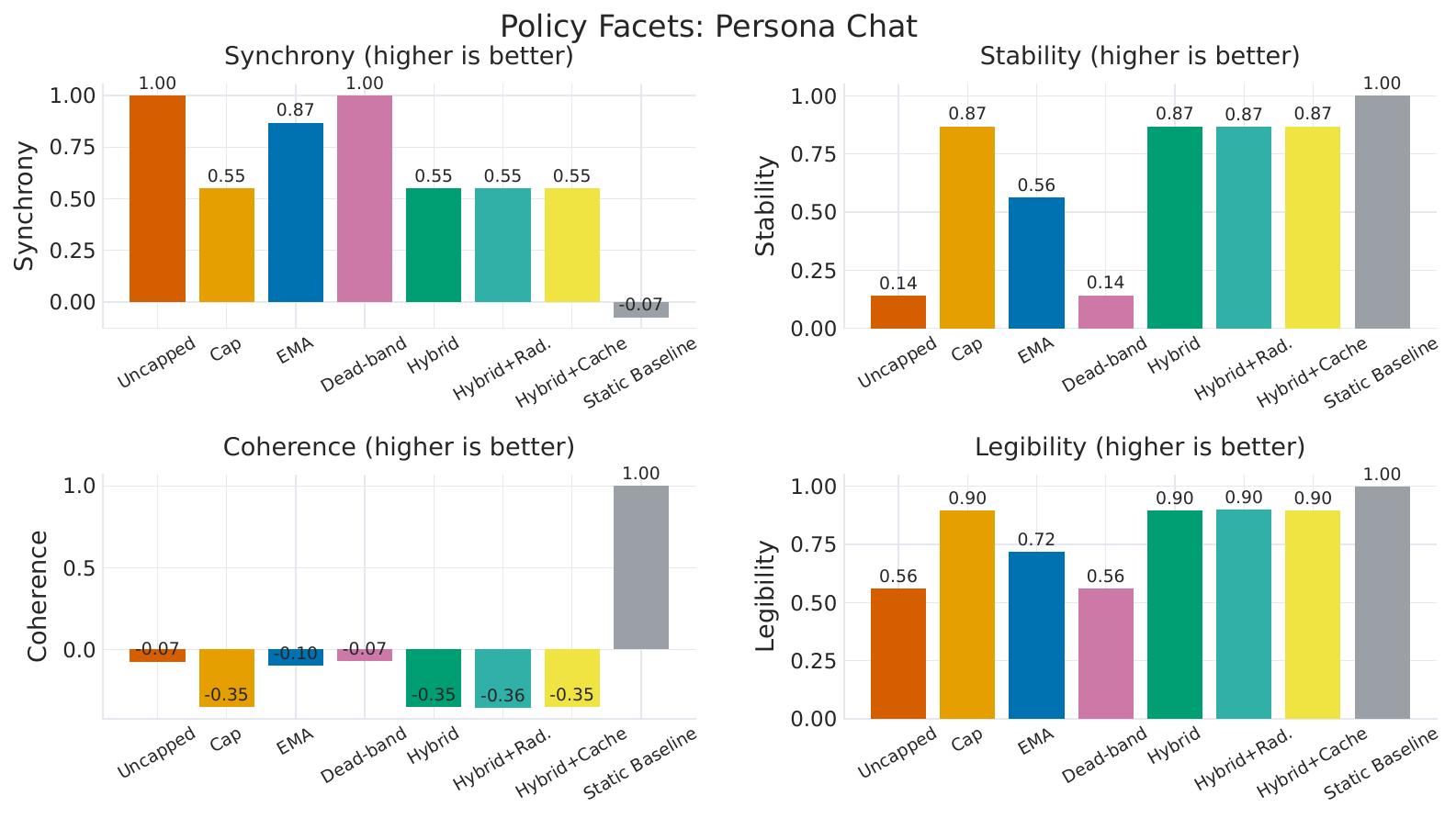}
  \vspace{1em}
  \includegraphics[width=0.71\linewidth]{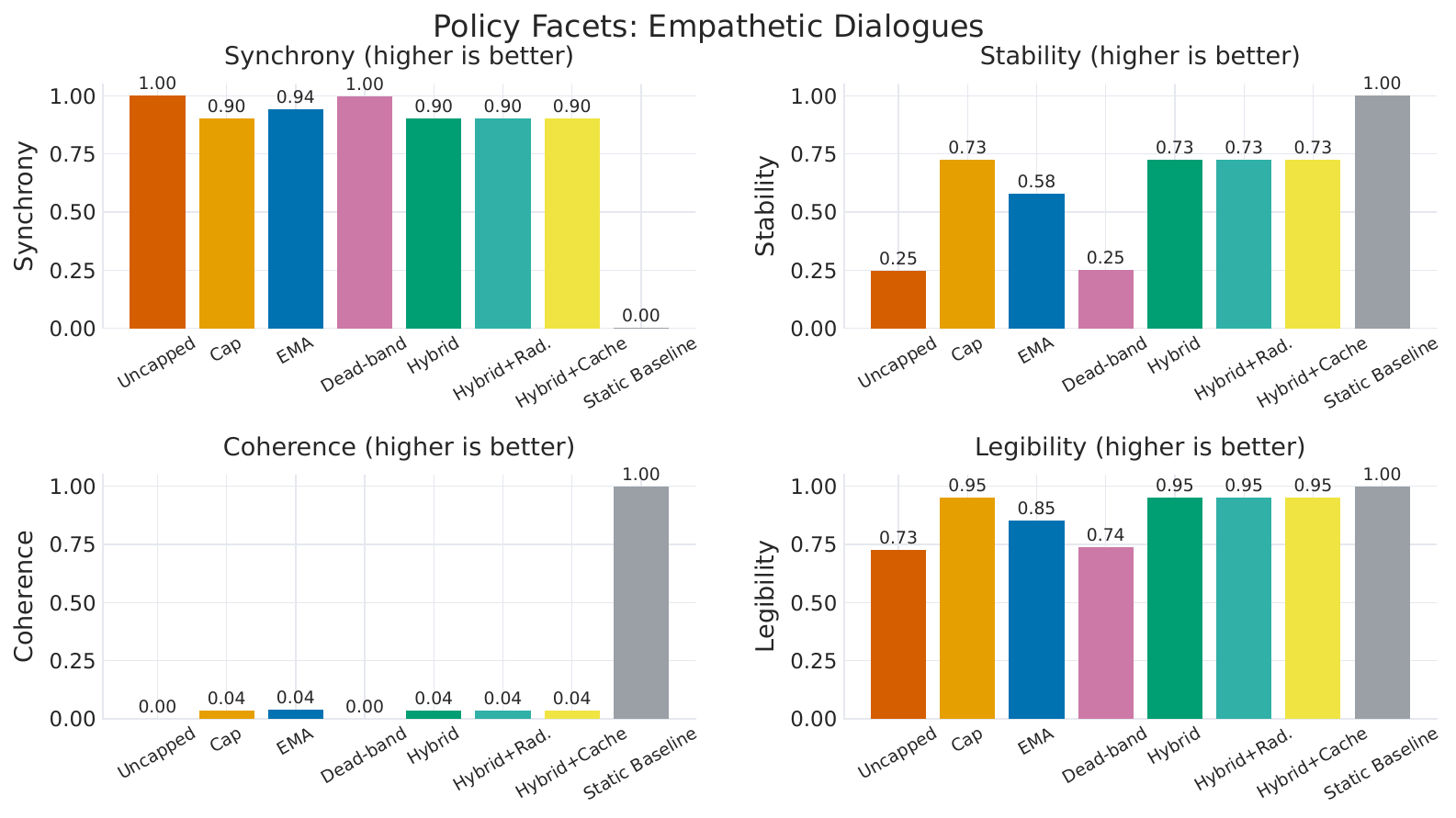}
  \caption{Policy Facets on External Datasets. Across all three corpora, bounded policies consistently improve \emph{Stability} relative to \emph{Uncapped}. The effect on \emph{Coherence} is context-dependent: on Persona-Chat, the stylistic mismatch between the average user and the assistant archetype is so large that highly stable policies (like Cap) result in lower coherence than pure mimicry.}
  \label{fig:external_facets}
\end{figure}
\FloatBarrier

\section{Per-Corpus Policy Summaries}
\label{sec:appendix_summaries}
To supplement the facet plots in Appendix~\ref{sec:appendix_dataset}, Table~\ref{tab:per-corpus-summaries} provides the numerical means for key metrics across the three external datasets.

\begin{table}[ht]
\centering
\small
\caption{{Per-corpus policy summaries (means).}}
\label{tab:per-corpus-summaries}
\begin{tabular}{lcccc}
\toprule
\textbf{Corpus / Policy} & \textbf{Synchrony} & \textbf{Stability} & \textbf{Coherence} & \textbf{Legibility} \\
\midrule
\multicolumn{5}{l}{\emph{DailyDialog}}\\
Uncapped & 1.000 & 0.183 & -0.010 & 0.658 \\
Dead-Band ($\epsilon{=}0.1$) & 0.997 & 0.187 & -0.009 & 0.663 \\
EMA ($\alpha$=0.5) & 0.909 & 0.569 & 0.004 & 0.798 \\
Cap (0.25) & 0.812 & 0.756 & -0.006 & 0.932 \\
Hybrid (EMA+Cap) & 0.812 & 0.756 & -0.005 & 0.932 \\
Hybrid+Radius & 0.812 & 0.756 & -0.006 & 0.933 \\
Static Baseline & -0.010 & 1.000 & 1.000 & 1.000 \\
\addlinespace
\multicolumn{5}{l}{\emph{Persona-Chat}}\\
Uncapped & 1.000 & 0.140 & -0.074 & 0.560 \\
Dead-Band ($\epsilon{=}0.1$) & 0.998 & 0.143 & -0.074 & 0.562 \\
EMA ($\alpha$=0.5) & 0.868 & 0.562 & -0.100 & 0.718 \\
Cap (0.25) & 0.550 & 0.869 & -0.354 & 0.897 \\
Hybrid (EMA+Cap) & 0.550 & 0.869 & -0.354 & 0.897 \\
Hybrid+Radius & 0.550 & 0.869 & -0.357 & 0.899 \\
Static Baseline & -0.074 & 1.000 & 1.000 & 1.000 \\
\addlinespace
\multicolumn{5}{l}{\emph{EmpatheticDialogues}}\\
Uncapped & 1.000 & 0.247 & 0.001 & 0.727 \\
Dead-Band ($\epsilon{=}0.1$) & 0.998 & 0.250 & 0.001 & 0.738 \\
EMA ($\alpha$=0.5) & 0.944 & 0.581 & 0.039 & 0.853 \\
Cap (0.25) & 0.903 & 0.725 & 0.035 & 0.950 \\
Hybrid (EMA+Cap) & 0.903 & 0.726 & 0.035 & 0.950 \\
Hybrid+Radius & 0.903 & 0.726 & 0.035 & 0.950 \\
Static Baseline & 0.001 & 1.000 & 1.000 & 1.000 \\
\bottomrule
\end{tabular}
\end{table}

\section*{GenAI Usage Disclosure}
Generative AI was used for LLM-in-the-loop simulations with OpenAI GPT-4.1 nano and Anthropic Claude Sonnet 4 (see Methods).

\bibliographystyle{plainnat}
\bibliography{references}

\end{document}